\documentclass[a4paper,final,conference]{IEEEtran}
\usepackage{mathpple}
\usepackage{times}

\usepackage[top=0.55in, bottom=0.55in, left=0.55in, right=0.55in]{geometry}

\usepackage{amsmath}  % Define \boldsymbol (in amsbsy too) and align
\usepackage{amssymb}  % Define \mathbb (in amsfonts too)
\usepackage{mathrsfs} % Define \mathscr, a script font

\usepackage{theorem}  % Helps in rendering theorems, etc.
\usepackage{cite}     % Gives multiple references as intervals
\usepackage{comment}  % Comments out with \begin{comment} ... \end{comment}

\usepackage{upref}
\usepackage{amsfonts}

\usepackage{verbatim}

\usepackage[dvipsnames,usenames]{color}
\usepackage{enumerate}

\usepackage{graphicx}

\usepackage{latexsym}

\usepackage{color}

\usepackage{multicol}
\usepackage{hhline}

\usepackage{psfrag}

\usepackage{tikz}
\usepackage{pgfplots}

\usepackage{hhline}

\usepackage[noend]{algpseudocode}

\usepackage{xcolor,colortbl}

\usepackage{hyperref}

\newcommand{\be}[1]{\begin{equation}\label{#1}}
\newcommand{\ee}{\end{equation}}

\newcommand{\bc}{\begin{center}}
\newcommand{\ec}{\end{center}}

\newcommand{\qed}{\hfill$\Box$\\[1ex]}

\newcommand{\cC}{{\cal C}}

\newcommand{\cO}{{\cal O}}

\renewcommand{\le}{\leqslant}

\renewcommand{\ge}{\geqslant}
\renewcommand{\geq}{\geqslant}

\newcommand{\C}{\mathbb{C}}

\newcommand{\Cref}[1]{Co\-rol\-la\-ry\,\ref{#1}}

\theoremstyle{plain} \theorembodyfont{\normalfont\slshape}

\newtheorem{thm}{Theorem$\!$}
\newenvironment{theorem}{\begin{thm}\hspace*{-1ex}{\bf.}}{\end{thm}}

\newtheorem{prop}[thm]{Proposition$\!$}
\newenvironment{proposition}{\begin{prop}\hspace*{-1ex}{\bf.}}{\end{prop}}

\newtheorem{lem}[thm]{Lemma$\!$}
\newenvironment{lemma}{\begin{lem}\hspace*{-1ex}{\bf.}}{\end{lem}}

\newtheorem{cor}[thm]{Corollary$\!$}
\newenvironment{corollary}{\begin{cor}\hspace*{-1ex}{\bf.}}{\end{cor}}

\newtheorem{conj}[thm]{Conjecture$\!$}
\newenvironment{conjecture}{\begin{conj}\hspace*{-1ex}{\bf.}}{\end{conj}}

\newtheorem{prob}[thm]{Problem$\!$}

\newtheorem{defi}[thm]{Definition$\!$}
\newenvironment{definition}{\begin{defi}\hspace*{-1ex}{\bf.}}{\end{defi}}

\theorembodyfont{\normalfont}

\newtheorem{exam}{Example$\!$}
\newenvironment{example}{\begin{exam}\hspace*{-1ex}{\bf .}}{\qed\end{exam}}

\newtheorem{remrk}{Remark$\!$}

\newtheorem{alg}[thm]{Algorithm$\!$}

\definecolor{Codecolor}{named}{White}  %{Tan}
\newcommand{\Copen}{\mbox{\{\kern-5.50pt\{}}
\newcommand{\Cclose}{\mbox{\}\kern-5.50pt\}}}
\newcommand{\Cslash}{\mbox{$\backslash\kern-6.02pt\backslash$}}

\newcommand{\red}[1]{{#1}}
\newcommand{\blue}[1]{{#1}}
\newcommand{\green}[1]{{#1}}
\newcommand{\brown}[1]{{#1}}

\begin{document}

\title{\textbf{PIR Codes with Short Block Length}\vspace{0ex}}% \\ \begin{large}\today\end{large}}

\date{\today}
\author{
\IEEEauthorblockN{\textbf{Sascha Kurz}}
\IEEEauthorblockA{%Computer Science Department  \\
Universit\"at Bayreuth \\
Universit\"atsstr. 30, Bayreuth, 95447, Germany \\
{\it sascha.kurz@uni-bayreuth.de}\vspace{-2ex}}
\and
\IEEEauthorblockN{\textbf{Eitan Yaakobi}}
\IEEEauthorblockA{%Computer Science Department  \\
Technion - Israel Institute of Technology\\
Haifa 3200003, Israel \\
{\it yaakobi@cs.technion.ac.il}}\vspace{-2ex}
}\vspace{-2ex}
\maketitle

\thispagestyle{empty}

\begin{abstract}
In this work \emph{private information retrieval (PIR) codes} are studied. In a $k$-PIR code, $s$ information bits are encoded in such a way that every information bit 
has $k$ mutually disjoint recovery sets. The main problem under this paradigm is to minimize the number of encoded bits given the values of $s$ and $k$, where this 
value is denoted by $P(s,k)$. The main focus of this work is to analyze $P(s,k)$ for a large range of parameters of $s$ and $k$. In particular, we improve upon several of the existing results on this value. 

\end{abstract}

\section{Introduction}\label{sec:intro}

A $k$-\emph{private information retrieval} (\emph{$k$-PIR}) \emph{code} is a coding scheme which encodes by linear operation some $s$ information bits to $n$ encoded 
bits such that each information bit has $k$ mutually disjoint recovery sets. The main figure of merit when studying PIR codes is the length $n$ of the code, given the 
values of $s$ and $k$. Thus, the value $P(s,k)$ denotes the minimum value of $n$ for which a length-$n$ $k$-PIR code exists. 
% Since it is known that for all fixed $k$, $\lim_{n\rightarrow \infty} B(n,k)/n=\lim_{n\rightarrow \infty} P(n,k)/n=1$,~\cite{IKOS04}, we evaluate these codes 
% by their redundancy and define $r_B(n,k) \triangleq B(n,k)-n, r_P(n,k) \triangleq P(n,k)-n$. 

PIR codes are similar in their definition to \emph{locally repairable codes} (\emph{LRC}) \emph{with availability}~\cite{PHO13,RPDV14,WZL15}, however PIR codes do not 
impose any constraint on the size of the recovery sets as done for LRCs. In fact, these codes have more in common with \emph{one-step majority-logic decodable codes} 
that were studied a while ago by Massey~\cite{Massey} and later by Lin and others~\cite{LC04} for applications of fast decoding. The main difference is that one-step 
majority-logic decodable codes require that each symbol (both information and redundancy) will have multiple recovery sets.

The rest of this section is organized as follows. In Section~\ref{sec:prelim}, we formally define the codes studied in the paper, list some of the known previous results 
which are relevant to our work, and discuss several preliminary results. In Section~\ref{sec:basic}, it is shown to how to construct PIR codes by lengthening existing codes. 
\brown{Additionally we give a geometric construction using $s$-dimensional simplex codes as a starting point.} In Section~\ref{sec_ILP} we present a general linear programming 
formulation for PIR codes which provides lower bounds on the parameters of these codes and in many cases completely determine the value of $P(s,k)$. 
\brown{Coding theoretic methods are applied in Section~\ref{sec_dual_distance} in order to obtain a few more lower bounds and exact values.} We fully solve the cases where $s=4$ or $s=5$ and 
present further lower and upper bounds in Section~\ref{sec:results}. There we also summarize the best known lower and upper bounds for small parameters in Table~\ref{tab_best_known_bounds}.  

\section{Definitions, Previous Work, and Preliminaries}\label{sec:prelim}
\subsection{Definitions}
In this section we formally define the codes studied in this paper. A binary linear code of length $n$ and dimension $s$ will be denoted by $[n,s]$ or $[n,s,d]$, where 
$d$ denotes its minimum Hamming distance. The set $[n]$ denotes the set of integers $\{1,2,\ldots,n\}$. The binary field is denoted by $\mathbb{F}_2$.

In this work we focus on \emph{private information retrieval} (\emph{PIR}) codes that were defined recently in~\cite{fazeli2015private}. This family of codes requires to 
encode some $s$ information bits into $n$ encoded bits such that every information bit has $k$ mutually disjoint recovery sets. Formally, these codes are defined as follows. 
\begin{definition}
An $[n,s]$ binary linear code $\cC$ will be called a \textbf{$k$-PIR code}, and will be denoted by $[n,s,k]_P$, if for every information bit $e_i, i \in [s]$, there exist 
$k$ mutually disjoint sets $R_{i,1},\ldots, R_{i,k}\subseteq [n]$ such that $e_i$ is the sum of the bits in $R_{i,j}$.
\end{definition}

The main problem in studying PIR codes is to minimize the length $n$ given the values of $s$ and $k$. We denote by $P(s,k)$ the smallest $n$ such that there exists an 
$[n,s,k]_P$ code and the optimal redundancy of $k$-PIR codes is defined by $r_P(s,k) \triangleq P(s,k)-s$. In case $k=1$, the code $[s,s]$ which simply stores all the 
information symbols is an $[s,s,1]_P$ PIR code, so that $P(s,1)=s$. Similarly, the simple parity check code $[s+1,s]$ is an $[s+1,s,2]_P$ PIR code which implies that $P(s,2)=s+1$.

% In~\cite{IKOS04}, it was shown using the subcube construction that for any fixed $k$ there exists an asymptotically optimal construction of $[N,n,k]_q^B$ batch code, and 
% hence $$\lim_{n\rightarrow \infty} B(n,k)_q/n=\lim_{n\rightarrow \infty} P(n,k)_q/n=1.$$ Therefore, it is important to study how fast the rate of these codes converges to one, 
% and so the redundancy of PIR and batch codes is studied. We define $r_B(n,k)_q$ to be the value $r_B(n,k)_q \triangleq B(n,k)_q-n$ and similarly, 
% $r_P(n,k)_q \triangleq P(n,k)_q-n$. Hence, $r_B(n,1)_q=r_P(n,1)_q =0$, $r_B(n,2)_q=r_P(n,2)_q=1$. 

\subsection{Previous Work}
In~\cite{fazeli2015private}, it was shown that for any fixed $k\geq3$ it is possible to construct $[n,s,k]$ PIR codes where $n=s+\cO(\sqrt{n})$, so 
$r_P(s,k)=\cO(\sqrt{s})$ for any fixed $k\geq3$, and in~\cite{rao2016lower,wootters2016linear} it was proved that $r_P(s,3)=\Omega(\sqrt{s})$. Since 
$r_P(n,k)\geq r_P(n,3)$ for any fixed $k\geq3$, these results assure also that for any fixed $k\geq 3$, $r_P(n,k)=\Theta(\sqrt{n})$. 

There are several known results and constructions of PIR codes; see e.g.~\cite{FGW17,lin2017lengthening,vajha2017binary}. 
We summarize here the most relevant known results for our problem: 
\begin{theorem}\label{thm_known_results}
\begin{enumerate}
\item $P(s_1+s_2,k)\le P(s_1,k)+P(s_2,k)$.
\item $P(s,k_1+k_2)\le P(s,k_1)+P(s,k_2)$.
\item $P(s,2k) = P(2,sk-1)+1$.
\item $r_P(s,k)= \Theta(\sqrt s)$ for fixed $k$,~\cite{fazeli2015private,rao2016lower,wootters2016linear}.
%\item $r_P(s,\sqrt{s})= \cO(s^{\frac{\log3}{2}})$,~\cite{LC04}.
%\item $r_P(s,s^{\epsilon})= \cO(s^{0.5+\epsilon})$,~\cite{LC04}.
%\item $r_P(s,s^{0.25})= \cO(s^{0.714})$,~\cite{FGW17}.
\item $P(s,2^{s-1}) = 2^s-1$. 
%\item For every integer $k\ge 1$ we have $P(1,k)=k$.
\item For every integer $k\ge 1$ we have $P(2,k)=\lceil 3k/2\rceil$.
\item For every even integer $k\ge 2$ we have $P(3,k)=\lceil 7k/4\rceil$.
\item For every positive $s$ and $k$, $P(s,k)\geq \frac{2^s-1}{2^{s-1}}k$ and equality if and only if $k$ is a multiple of $2^{s-1}$, \cite[Theorem 16]{fazeli2015private})
\end{enumerate}
\end{theorem}

%% \begin{comment}
%% If the dimension $s$ is at most $3$, the exact values of $P(s,k)$ are also well known.
%% \begin{proposition}
%%   \label{prop_dimension_at_most_3_exact}
%%   \begin{enumerate}
%%     \item[(a)] For every integer $k\ge 1$ we have $P(1,k)=k$.
%%     \item[(b)] For every integer $k\ge 1$ we have $P(2,k)=\lceil 3k/2\rceil$.
%%     \item[(c)] For every even integer $k\ge 2$ we have $P(3,k)=\lceil 7k/4\rceil$.
%%   \end{enumerate}  
%% \end{proposition}
%% \end{comment}

% Our goal in this work is to close on this gap and study new constructions of PIR and batch codes. Lastly, we note that we only focus on the case where $n$ is large. 
% The case of fixed $n$ or small $n$ comparing to $k$ was studied in~\cite{FVY15,LR17,vajha2017binary} for PIR codes.

% Yet another class of related codes is called \emph{combinatorial batch codes}. Here, it is required to have the same properties of batch codes, however the symbols 
% cannot be encoded and thus these codes cannot be compared with the ones studied in this paper. Several works have considered codes under this setup; see e.g.~\cite{BRR12,BKMS10,BT12,SG13}.

Several more construction of PIR codes from bipartite graphs and constant weight codes were constructed in \cite[Section IV.D]{fazeli2015private}. %%\footnote{The function $A$ in Theorem~10 is an abbreviation for $P(s,k)$.} 
One-step majority logic codes where used to obtain the following result. 
\begin{theorem}(\cite[Theorem 9]{7282977},\cite[Theorem 8]{fazeli2015private})\\
  For any $\theta$, $\ell$, and $\lambda$ we have
  \begin{eqnarray*}
    P\left(2^{2\theta \ell}-\left(2^{\theta+1}-1\right)^\ell-1,2^\ell+2\right) &\le& 2^{2\theta \ell}-1.\\
    P\left(\left(2^\lambda-1\right)^\ell-2,2^\ell\right)&\le&2^{\lambda \ell}-1.
  \end{eqnarray*}
\end{theorem}
We remark that the formulations of \cite[Theorem 9]{7282977} and \cite[Theorem 8]{fazeli2015private} have been slightly adjusted by reducing the value of $s$ by one. As already observed in \cite{lin2017lengthening}, applying $\theta=1$, $\ell=2$ in the first formula gives $P(7,6)\le 15$. \red{However, $P(7,6)\ge N(7,6)=16$, see (\ref{ie_coding_theoretic_lower_bound}), which is a contradiction.} In \cite[Section V, Remark 3]{lin2017lengthening} the authors 
traced back the problem back to a misprint in \cite[p. 289]{costello1982error}, see \cite[Theorem 6]{lin1980class} for the correct version. 

An analytic solution of the general lengthening problem, see the discussion in Section~\ref{sec_preliminaries}, resulted with the following.
\begin{theorem}\label{th:length}
(\cite[Theorem 1]{lin2017lengthening}) 
$P(s+1,k)\le P(s,k)+\left\lceil\frac{k}{2}\right\rceil$.
\end{theorem}
%\begin{IEEEproof} Assume $k$ even and conclude the result for odd $k$ by the parity-check-bit argument. Let $\mathcal{R}^i$ be the recovery sets for $e_i$, where $1\le i\le s$, of a generator matrix $G$ of an $s$-dimensional $k$-PIR code of length $n$. Set $t=k/2$ and let $p_1,\dots,p_t\in \{1,\dots,n\}$ that are contained in exactly $t$ elements $R_1^1,\dots R_t^1$ of $\mathcal{R}^1$. With $r\in\mathbb{F}_2^n$ being the vector that has its ones exactly in the positions $p_1,\dots, p_t$ we construct a lengthened generator matrix $G'$ from $G$ by appending row $r$ and $t$ times the column $e_{s+1}$. For each $1\le i\le s$ at most $t$ elements of $\mathcal{R}^i$ can contain an odd number of ones in $r$ (since there are only $t$ ones), so that we can use one of the $t$ additional columns $e_{s+1}$ to construct $k$ recovers sets for each $e_i$ (with $1\le i\le s$) in $G'$. For unit vector $e_{s+1}$ we pair $R_j^1$ with one recovery set from $\mathcal{R}^1\backslash\{R_1^1,\dots R_t^1\}$ so that we obtain a partition of $\mathcal{R}^1$. By construction these sets are $t$ recovery sets for $e_{s+1}$ in $G'$. The missing $t$ recovery sets are given by the $t$ columns $e_{s+1}$. \end{IEEEproof}

Since the minimum Hamming distance of every $k$-PIR code is at least $k$ it holds that 
\begin{equation}\label{ie_coding_theoretic_lower_bound}
P(s,k)\ge N(s,k),
\end{equation}   
where $N(s,k)$ denotes the the smallest integer $n$ such that a $[n,s,k]$ code exists. 
\blue{In \cite{griesmer1960bound} the so-called Griesmer bound 
$$  N(s,k)\ge \sum_{i=0}^{s-1} \left\lceil\frac{k}{2^i}\right\rceil=:G(s,k) $$ 
was proven. Interestingly enough, for every fix integer $s$ we have $N(s,k)=G(s,k)$ if $k$ is sufficiently large \cite{baumert1972note}, i.e., for 
every fix integer $s$ the determination of the function $N(s,k)$ is a finite problem. Those functions are explicitly known for all $s\le 8$ 
\cite{bouyukhev2000smallest}.} Much research has been devoted on the determination of $N(s,k)$ for specific parameters $s$ and $k$. For the currently best known lower and upper bounds on $N(s,k)$ we refer to the online tables \cite{Grassl:codetables}.\footnote{We remark that 
with respect to lower bounds on $N(s,k)$ it makes also sense to check the entries at \url{http://mint.sbg.ac.at} which sometimes contain improvements.} The values of $N(s,k)$ are a good benchmark for constructions of $s$-dimensional binary $k$-PIR codes, i.e., constructive upper bounds for $P(s,k)$. If $N(s,k)$ is met, then the corresponding PIR code is obviously optimal. 

It is quite hard to get better lower bounds than $P(s,k)\ge N(s,k)$. One parametric improvement was stated in the literature so far: $P(s,3)\ge s+\left\lceil\sqrt{2s+\frac{1}{4}}+\frac{1}{2}\right\rceil$, see \cite[Theorem 3, Equation 10]{rao2016lower}. If 
we combine it with the puncturing constraint $P(s,k)\ge P(s,k-1)+1$, then we obtain
\begin{equation}
  P(s,k)\ge \left\lceil\sqrt{2s+\frac{1}{4}}+\frac{1}{2}\right\rceil+k-3
  \label{ie_lower_bound_vardy_rao}
\end{equation}
for $k\ge 3$. We remark that for $k=3$ or $k=4$ this inequality is always at least as good as the coding theoretic lower bound $P(s,k)\ge N(s,k)$ and it is indeed a 
strict improvement for larger values of $s$.\footnote{\label{fn_rao_bound_k_3}More precisely, it is a strict improvement for $s=4$ and all $s\ge 7$. An exact formula for $N(s,3)$, and so 
also for $N(s,4)$, exists. It is attained by the Hamming codes and puncturings thereof. The lower bound follows from the Hamming or sphere packing bound.} For systematic PIR codes, see 
the explanation below, the same bound was also proved in \cite{wootters2016linear} and \cite{vajha2017binary}.

Despite significant progress on determining the exact value of $P(s,k)$, this problem is far from being solved. The goal of this paper is to build upon previous work and develop new tools which are specifically targeted towards deriving upper and lower bound on $P(s,k)$ in order to establish many cases which still remained open. For example, if we apply Theorem~\ref{th:length} to a $9$-dimensional $10$-PIR code of length $28$, we can conclude $P(10,10)\le 33$, which improved the best known construction. By solving an instance of the general lengthening problem this can even be improved to $P(10,10)\le 31$, see the discussion in Section~\ref{sec_preliminaries}.  

%\begin{center} \textbf{ToDo:} The constructions based on Steiner systems, see \cite[Section III.B]{lin2017lengthening} and \cite[Section IV.B]{fazeli2015private}, look  very interesting and the search problems might be easily formulated as ILP problems. \end{center}  

%\begin{center} \textbf{ToDo:} The general construction of binary shortened projective Reed-Muller codes seems to be very strong, see \cite{vajha2017binary} for the details and the numerical bounds for the best known codes in Table~\ref{tab_best_known_bounds}. The attained demands $k$ are powers of $2$. Further related possibly interesting papers are \cite{carvalho2019projective,ramkumar2018determining}.\end{center}

\subsection{Preliminaries}\label{sec_preliminaries}
Note that an $[n,s]$ code is a $k$-PIR code if it admits a generator matrix $G\in \mathbb{F}_2^{s\times n}$ such that for each $1\le i\le s$ there exist disjoint sets 
$R_1^i,\dots,R_k^i\subseteq [n]$ such that $\sum_{h\in R_j^i} G^h=e_i$ for all $1\le j\le k$, where $G^h$ denotes the $h$th column of $G$ and $e_i$ denotes the $i$th 
unit vector. The interpretation is that $e_i$ can be recovered by the $k$ disjoint sets $R_j^i$, which are therefore also called \emph{recovery sets}. The set of all 
recovery sets for $e_i$ is denoted by $\mathcal{R}_i$, i.e., $\mathcal{R}_i=\left\{ R_j^i \mid 1\le j\le k\right\}$. %%\cup_j R_j^i$. 
We call a recovery set $R$ for $e_i$ \emph{minimal} if no proper subset $R'\subsetneq R$ is a recovery set for $e_i$.  
%Note that given $G$ there can be different lists of recovers sets, see Example~\ref{example_generator_matrix_recovery_sets}. So, even if $P(s,k)$ is attained, we may optimize further parameters. The largest size of a used recovery set might be an interesting, possibly practical relevant such parameter. In some applications we use different representations for recovery sets, see also Example~\ref{example_generator_matrix_recovery_sets}. We may speak of the \textit{label} or \textit{point representation} (\textit{vector representation}?).

\begin{example}
\label{example_generator_matrix_recovery_sets}
An example for a generator matrix attaining $P(4,4)=9$ is given by: 
$$
  G=\begin{pmatrix}
    1 0 0 0 1 1 1 1 1\\ 
    0 1 0 0 0 1 0 1 1\\ 
    0 0 1 0 1 1 0 0 1\\ 
    0 0 0 1 0 0 1 1 1\\
  \end{pmatrix}.
%%  \quad
%%  G'=\begin{pmatrix}
%%    1 0 0 0 1 1 1 1 1\\ 
%%    0 1 0 0 0 1 0 1 1\\ 
%%    0 0 1 0 1 1 0 0 1\\ 
%%  \end{pmatrix}.
$$
For $e_4$ we can use the recovery sets
$$
  \{4\}, \{1,7\}, \{6,9\}, \{2,3,5,8\}.
$$    
Note that there is also a different list of recovery sets:
$$
  \{4\}, \{6,9\}, \{3,5,7\}, \{1,2,8\}.
$$
The later might have the advantage that it only uses recovery sets of cardinality at most $3$. As a different notation for recovery sets we also 
use the columns directly (instead of their labels). In our last example $\mathcal{R}_4$ then reads\vspace{-2ex}

\begin{small}
\begin{eqnarray*}
  \left\{\hspace{-1ex}\begin{pmatrix}0\\0\\0\\1\end{pmatrix}\hspace{-1ex}\right\},
  \left\{\hspace{-1ex}\begin{pmatrix}1\\1\\1\\0\end{pmatrix},\hspace{-0.5ex}\begin{pmatrix}1\\1\\1\\1\end{pmatrix}\hspace{-1ex}\right\},
  \left\{\hspace{-1ex}\begin{pmatrix}0\\0\\1\\0\end{pmatrix},\hspace{-0.5ex}\begin{pmatrix}1\\0\\1\\0\end{pmatrix},\hspace{-0.5ex}\begin{pmatrix}1\\0\\0\\1\end{pmatrix}\hspace{-1ex}\right\},
  \left\{\hspace{-1ex}\begin{pmatrix}1\\0\\0\\0\end{pmatrix},\hspace{-0.5ex}\begin{pmatrix}0\\1\\0\\0\end{pmatrix},\hspace{-0.5ex}\begin{pmatrix}1\\1\\0\\1\end{pmatrix}\hspace{-1ex}\right\}.
\end{eqnarray*}
\end{small}
For minimal recovery sets these are indeed sets of non-zero vectors in $\mathbb{F}_2^s$ and multisets in general. In the latter case points 
can have a multiplicity larger than $1$. 
\end{example} 

\blue{Each binary linear $[n,s]$ code $C$  of effective length $n$ is in bijection to a multiset $\mathcal{P}$ of points, i.e., $1$-dimensional subspaces of $\mathbb{F}_2^s$, of cardinality $n$. 
Starting from a generator matrix $G$ we can obtain a multiset of points by choosing the points $\langle G^k\rangle$ for every column $G$. In the other direction we can choose an arbitrary 
generator for each point, i.e., a column vector, and build up a generator matrix with those column vectors. This geometrical point of view can gives an easy way to form constructions of PIR codes. For example, by taking the set of all $2^s-1$ points in $\mathbb{F}_2^s$, we get the so-called \emph{$s$-dimensional simplex code} in order to get the known result of $P(s,2^{s-1})\le 2^s-1$ for all $s\ge 1$.\footnote{As an abbreviation set $k=2^{s-1}$ and number the $2^{s-1}$ vectors of $\mathbb{F}_2^s$ with $i$th component equal to 
 zero by $x^{1,i},\dots, x^{k,i}$, where we assume that $x^{1,i}$ is the zero vector. For each $1\le i\le s$ we can take the recovery sets $R_1^i=\left\{x^{1,i}+e_i\right\}$ and 
 $R_j^i=\left\{x^{j,i},x^{j,i}+e_i\right\}$ for $2\le j\le k$.\label{fn_simplex_code}} } We will use this geometric formulation when deriving several of our results, see 
 especially Proposition~\ref{prop_remove_lines}.

\blue{For a lower bound let $H$ be an arbitrary hyperplane of $\mathbb{F}_2^s$. Since $\dim(H)=s-1$ there exists at least one index $1\le i\le s$ such that 
$\left\langle e_i\right\rangle $ is not contained in $H$. Thus, there have to be at least $k$ points outside of $H$ to form the recovery sets:
\begin{lemma}(C.f.\ \cite[Theorem 2]{skachek2018batch}, \cite[Lemma 2]{lipmaa2015linear})\\
  Let $\mathcal{P}$ be the multiset of points corresponding to an $s$-dimensional $k$-PIR. For every hyperplane $H$ of $\mathbb{F}_2^s$ we have 
  \begin{equation}
    \label{ie_hyperplane_argument}
    | \left\{ P\in\mathcal{P} \mid P\notin H \right\}|\ge k,
  \end{equation}
  counting points with there respective multiplicity.\label{lemma_hyperplane_argument}
\end{lemma} 
Summing up Inequality~(\ref{ie_hyperplane_argument}) for the $2^s-1$ hyperplanes gives $P(s,k)\ge \frac{2^s-1}{2^{s-1}}\cdot k$, see \cite[Theorem 16]{fazeli2015private}, 
since each point of $\mathbb{F}_2^s$ is contained in exactly $2^{s-1}-1$ hyperplanes. While this gives $P\left(s,2^{s-1}\right)=2^s-1$, it can be improved easily. 
Since each hyperplane of $\mathbb{F}_2^s$ corresponds to a codeword $c_H$ of the code whose weight equals the number of points outside of $H$, we have 
the well known fact that each $k$-PIR code has a minimum Hamming distance of at least $k$. Thus, we have
%% Since the minimum Hamming distance of every $k$-PIR code is at least $k$ it holds that 
\begin{equation}
  P(s,k)\ge N(s,k),
  \label{ie_coding_theoretic_lower_bound}
\end{equation}   
where $N(s,k)$ denotes the the smallest integer $n$ such that a $[n,s,k]$ code exists. In \cite{griesmer1960bound} the so-called Griesmer bound 
$$  N(s,k)\ge \sum_{i=0}^{s-1} \left\lceil\frac{k}{2^i}\right\rceil=:G(s,k) $$ 
was proven. Interestingly enough, for every fix integer $s$ we have $N(s,k)=G(s,k)$ if $k$ is sufficiently large \cite{baumert1972note}, i.e., for 
every fix integer $s$ the determination of the function $N(s,k)$ is a finite problem. Those functions are explicitly known for all $s\le 8$ 
\cite{bouyukhev2000smallest}. Much research has been devoted on the determination of $N(s,k)$ for specific parameters $s$ and $k$. For the currently best known lower and upper bounds on $N(s,k)$ we refer to the online tables \cite{Grassl:codetables}.\footnote{We remark that 
with respect to lower bounds on $N(s,k)$ it makes also sense to check the entries at \url{http://mint.sbg.ac.at} which sometimes contain improvements.} The values of $N(s,k)$ are a good benchmark for constructions of $s$-dimensional binary $k$-PIR codes, i.e., constructive upper bounds for $P(s,k)$. If $N(s,k)$ is met, then the corresponding PIR code is obviously optimal. }

\blue{It is quite hard to get better lower bounds than $P(s,k)\ge N(s,k)$. One parametric improvement was stated in the literature so far: $P(s,3)\ge s+\left\lceil\sqrt{2s+\frac{1}{4}}+\frac{1}{2}\right\rceil$, see \cite[Theorem 3, Equation 10]{rao2016lower}. If 
we combine it with the puncturing constraint $P(s,k)\ge P(s,k-1)+1$, then we obtain
\begin{equation}
  P(s,k)\ge \left\lceil\sqrt{2s+\frac{1}{4}}+\frac{1}{2}\right\rceil+k-3
  \label{ie_lower_bound_vardy_rao}
\end{equation}
for $k\ge 3$. We remark that for $k=3$ or $k=4$ this inequality is always at least as good as the coding theoretic lower bound $P(s,k)\ge N(s,k)$ and it is indeed a 
strict improvement for larger values of $s$.\footnote{\label{fn_rao_bound_k_3}More precisely, it is a strict improvement for $s=4$ and all $s\ge 7$. An exact formula for $N(s,3)$, and so 
also for $N(s,4)$, exists. It is attained by the Hamming codes and puncturings thereof. The lower bound follows from the Hamming or sphere packing bound.} For systematic PIR codes, see 
the explanation below, the same bound was also proved in \cite{wootters2016linear} and \cite{vajha2017binary}.}

We call a generator matrix $G$ of a linear code systematic if it contains a unit matrix. While every linear code admits a systematic generator 
matrix it is not clear whether there always exists a systematic PIR code matching $P(s,k)$, see Question~4 in \cite[Sec. 10]{skachek2018batch}. 
Here we give an example that this is not the case. More specifically, we will show $P(6,8)=19$ in Section~\ref{sec_ILP} while every systematic PIR code has length at least 
$20$, see Proposition~\ref{prop_non_systematic_is_better}. %%(and there indeed exists an example of that length). 
An optimal non-systematic generator matrix (of length $19$) is given by 
$$
  %%\left(\begin{smallmatrix}
  \begin{pmatrix}
    0 1 0 1 1 1 1 0 1 1 1 1 0 1 1 1 0 1 1\\ 
    1 1 1 1 0 1 0 1 1 1 1 0 1 1 0 1 1 1 0\\ 
    1 1 1 1 0 0 1 1 1 1 0 1 1 1 1 1 0 0 1\\ 
    1 1 0 0 1 1 1 1 1 0 1 1 1 1 0 0 1 1 1\\ 
    1 1 0 0 0 0 0 0 0 1 1 1 1 1 1 1 1 1 1\\ 
    0 0 1 1 1 1 1 1 1 1 1 1 1 1 0 0 0 0 0\\ 
  \end{pmatrix}
  %%\end{smallmatrix}\right)
$$ 
and an optimal systematic generator matrix (of length $20$) is given by
$$
  \begin{pmatrix}
    1 0 0 0 0 0 0 1 1 1 1 1 1 1 1 1 0 0 1 1\\ 
    0 1 0 0 0 0 0 1 0 1 1 0 1 1 0 0 1 1 1 1\\ 
    0 0 1 0 0 0 1 1 0 0 1 1 1 1 1 0 0 0 0 0\\ 
    0 0 0 1 0 0 1 1 1 1 1 0 0 1 0 1 0 0 0 0\\ 
    0 0 0 0 1 0 0 0 1 1 1 0 0 0 1 1 1 1 1 1\\ 
    0 0 0 0 0 1 0 0 0 0 0 1 1 1 1 1 1 1 1 1\\ 
  \end{pmatrix}.
$$

We call a code projective if all columns of an arbitrary generator matrix are pairwise linear independent, i.e., no column is a multiple of another one. 
Note that the last-but-one generator matrix (and so the code) is projective, while the last generator matrix is not, since the last four columns correspond to 
two points with multiplicity $2$ each.

According to this observation we notice that while the objects under consideration are called PIR codes, in fact their properties actually can depend on their generator matrix. This is different for codes with disjoint repair groups as e.g.\ studied in \cite{wootters2016linear}. The code property of disjoint repair 
groups is more demanding than that of PIR codes (depending on the generator matrix). For systematic generator matrices both notions are the same, so that the lower bound from \cite{wootters2016linear} only works (directly) for systematic generator matrices.

\blue{Let $\mathcal{P}$ be a multiset of points in $\mathbb{F}_2^s$ and $n$ denote its cardinality $\#\mathcal{P}$. By $h_i$ we denote the number of hyperplanes of $\mathbb{F}_2^s$ that contain exactly $i$ points. Counting incidences $(H)$, $(H,P)$, and $(H,P,P')$, where the $H$ are hyperplanes and the $P,P'$ are different points in $\mathcal{P}$, gives the so-called standard equations
\begin{eqnarray}
  \sum_{i\ge 0} h_i       &=& 2^s-1 \label{se1}\\
  \sum_{i\ge 0} ih_i      &=& n\left(2^{s-1}-1\right) \label{se2}\\
  \sum_{i\ge 0} i(i-1)h_i &=& n(n-1)\left(2^{s-2}-1\right)+2^{s-2}y_2 \label{se3},
\end{eqnarray} 
where $y_2$ denotes the number of ordered different pairs of $\mathcal{P}$ that correspond to the same geometric point.} 
\brown{We remark that the above so-called standard equations are a geometric variant of the first 3 MacWilliams identities. We will use them in the proof of Lemma~\ref{lemma_reed_muller_unique}.}

\section{Basic Constructions of PIR Codes}\label{sec:basic}
Next we consider how to lengthen a $k$-PIR code in order to increase the number of information bits it stores and still preserve its property as a $k$-PIR code. Hence, we add columns and a row to a generator matrix, where we consider the special case, where the new columns all are unit vectors with the one entry in the last row.
\begin{proposition}
  \label{prop_lengthening}
  Let $G$ be a generator matrix of an $s$-dimensional linear code of length $n$. If $G$ can be lengthened by one row and $t$ columns of $e_{s+1}$ to 
  a generator matrix $G'$ of an $(s+1)$-dimensional $k$-PIR code, then
  \begin{enumerate} 
    \item[(1)] $G'$ has length $n+t$,
    \item[(2)] $G$ is a $k$-PIR code,
    \item[(3)] for all $1\le i\le s$ there exist recovery sets $\mathcal{R}^i$ of $e_i$ in $G$ such that for $G'$ the recovery sets each yield 
               either $e_i$ or $e_i+e_{s+1}$, and the latter case occurs at most $t$ times,
    \item[(4)] the recovery sets of $e_{s+1}$ in $G'$ without those that only consist of a unit vectors $e_{s+1}$ have the property that for $G$ they sum
               to zero and there are at least $k-t$ of them.            
  \end{enumerate}   
\end{proposition}
\begin{IEEEproof}
  The length of $G'$ directly follows from the construction, so that (1) holds. Let $\mathcal{R}'^i$ be the recovery sets of $e_i$ in $G'$ for $1\le i\le s$. Over $G'$ one of 
  these recovery sets sums to $e_i$. W.l.o.g.\ we assume that the recovery sets are reduced so that they contain $e_{s+1}$ at most once. If $e_{s+1}$ is 
  not contained in the recovery set, then it is also a recovery set for $e_i$ in $G$. If $e_{s+1}$ is contained in the recovery set, then we can remove 
  $e_{s+1}$ and obtain a recovery set of $e_i$ in $G$ that sums to $e_i+e_s$ in $G'$. Of course the latter case can occur at most $t$ times since the $e_{s+1}$ is 
  contained exactly $t$ times as a column in $G'$. This gives (3) and (2). Consider the set $\mathcal{R}'^{s+1}$ of the $k$ recovery sets of $e_{s+1}$ in $G'$. At least 
  $k-t$ are not given by the singleton $\{e_{s+1}\}$. Since the corresponding columns in $G'$ sum to $e_{s+1}$, they sum to zero in $G$.
\end{IEEEproof}
 
The insights of Proposition~\ref{prop_lengthening} can be turned into the following algorithm to generate $(s+1)$-dimensional $k$-PIR codes of length $n+t$ from $s$-dimensional $k$-PIR codes of length $n$.
\begin{enumerate}
  \item[(1)] Compute a list $\mathcal{C}_1$ of recovery sets $\left(\mathcal{R}^i\right)_{1\le i\le s}$ of the unit vectors in $G$ such that $|\mathcal{R}^i|=k$.
  \item[(2)] Compute a list $\mathcal{C}_2$ of a set $\mathcal{Z}$ of disjoint sets whose corresponding columns in $G$ sum to zero, where $|\mathcal{Z}|\ge k-t$.
  \item[(3)] Loop over all $r\in\mathbb{F}_2^n$ such that there exists $\left(\mathcal{R}^i\right)_{1\le i\le s}$ in $\mathcal{C}_1$ so that for each 
             $1\le i\le s$ from the $k$ elements in $\mathcal{R}^i$ we have an odd number of $1$s in the corresponding positions in $r$ in at most $t$ cases.
  \item[(4)] If there exists an element $\mathcal{Z}$ in $\mathcal{C}_2$ such that from the elements in $\mathcal{Z}$ we have an odd number of $1$s in the 
             corresponding positions in $r$ in at least $k-t$ cases, $\begin{pmatrix} G & 0\\ r& 1\dots 1\end{pmatrix}$ is the generator matrix of an $(s+1)$-dimensional 
             $k$-PIR code of length $n+t$.           
\end{enumerate}
 
We remark that recovery sets as well as dual codewords might be found looping over all binary vectors of length $n$. Collections of disjoint recovery sets $\mathcal{R}^i$ 
or the disjoint zero sets $\mathcal{Z}$ might be found by clique search. Of course a promising heuristic is to check only binary vectors of rather small weight. 

\begin{example}
In this example, we consider the $9$-dimensional $10$-PIR code of length $P(9,10)=28$ with generator matrix
$$
   \begin{pmatrix}
     1 0 0 0 0 0 0 0 0 1 0 1 0 1 0 0 0 1 0 1 1 0 1 0 0 1 0 1\\ 
     0 1 0 0 0 0 0 0 0 1 1 1 0 0 1 0 1 1 0 0 0 0 1 0 1 0 0 1\\ 
     0 0 1 0 0 0 0 0 0 0 1 0 1 0 1 0 0 1 1 0 1 0 0 1 1 0 0 1\\ 
     0 0 0 1 0 0 0 0 0 0 0 1 1 0 0 1 0 0 1 1 1 0 0 0 1 0 1 1\\ 
     0 0 0 0 1 0 0 0 0 1 0 1 0 0 1 1 0 0 0 0 1 1 0 0 0 1 1 1\\ 
     0 0 0 0 0 1 0 0 0 0 1 1 0 1 1 0 0 0 1 0 0 1 1 0 0 1 0 1\\ 
     0 0 0 0 0 0 1 0 0 0 0 0 1 1 1 0 1 1 1 0 0 0 1 0 0 0 1 1\\ 
     0 0 0 0 0 0 0 1 0 0 0 0 0 0 0 1 1 1 1 0 0 0 0 1 1 1 1 1\\ 
     0 0 0 0 0 0 0 0 1 0 0 0 0 0 0 0 0 0 0 1 1 1 1 1 1 1 1 1\\   
   \end{pmatrix}
$$
and try $t=3$.  It comes with 
a collection of recovery sets with cardinality distribution \red{$1^1 3^9$} for each $1\le i\le s$, which we took as our single candidate in $\mathcal{C}_1$. 
For $\mathcal{C}_2$ we build up a graph with $524\, 287$~nodes and determined a clique of maximum cardinality $7$. After less than 2~minutes computation time 
we found the first extension $r=\begin{pmatrix}0 0 0 0 0 0 0 1 0 1 0 0 1 0 0 0 1 0 0 1 1 0 0 1 0 0 0 0 \end{pmatrix}$. For $e_{10}$ the recovery sets in label 
notation are given by $\{4,6,8,12\}$, $\{0,2,15,16\}$, $\{10,11,20,21\}$, $\{1,3,23,24\}$, $\{5,14,19,25\}$, $\{9,13,22,26\}$, and $\{7,17,18,27\}$. After less than 
six minutes we found $9$ different extensions certifying $P(10,10)\le 31$ in total. So, in our example we have chosen both $\mathcal{C}_1$ and $\mathcal{C}_2$ of cardinality 
$1$. 
\end{example}

In general, if we apply the above algorithm as a heuristic and not for exhaustive enumeration, we do not need to find all possibilities. As mentioned above, the same applies to the possibilities for the extension row $r$. This rough idea leaves a lot of space for algorithmic implementations.

%\begin{center} \textbf{ToDo:} In \cite{rao2016lower} the authors state that the proof in \cite{wootters2016linear} uses different techniques. However, for me, the proofs in \cite{rao2016lower}, \cite{wootters2016linear}, and \cite{vajha2017binary} are rather similar. \end{center}

\blue{It was stated in \cite{lin2017lengthening} that the coding theoretic lower bound is tighter if, given dimension $s$, the number $k$ of disjoint recovery sets is relatively small. Their formulation might be interpreted 
in the way that they claim that for $k=6$ (or equivalently $k=5$) the coding theoretic bound is always at least as tight as Inequality~(\ref{ie_lower_bound_vardy_rao}).\footnote{{\lq\lq}It 
can be easily verified that in general $N(k,t)\le L_P(k,t)$ for small values of $t>4$. In fact, we will show in Section~V that $N(k,t)$ is a tighter lower bound on $N_P(k,t)$ than 
$L_P(k,t)$ for $r=6$.{\rq\rq}.} However, this is not the case. An example is given by $s=92$ and $k=5$, where $N(92,5)=106$\footnote{The value 
$N(92,5)=106$ is taken from \url{http://mint.sbg.ac.at}.} but 
$P(92,5)\ge 107$ due to Inequality~(\ref{ie_lower_bound_vardy_rao}). Also for $t>6$ there are such examples, however, they require rather large values of $s$. So, the situation should be as follows: If $k>4$ and the dimension $s$ is not too big with respect to $k$, then the coding theoretic lower bound is superior. If the dimension gets \textit{huge}, then Inequality~(\ref{ie_lower_bound_vardy_rao}) is tighter. For $k=3$ or $k=4$, see Footnote~\ref{fn_rao_bound_k_3}. }

Using even more geometric terms, we can formulate a parametric construction. The rough idea is a follows. We start with an $s$-dimensional simplex code, 
represented by a set of points $\mathcal{P}$, and the recovery sets stated in Footnote~\ref{fn_simplex_code}. A \emph{line} in $\mathbb{F}_2^s$ is a set 
of three collinear points, i.e., three non-zero vectors $a,b,c$ in $\mathbb{F}_2^s$ with $a+b+c=\mathbf{0}$. We iteratively remove the three points 
from a line from $\mathcal{P}$ and modify the list $\mathcal{R}^i$ of recovery sets accordingly. A \emph{partial line spread} in $\mathbb{F}_2^s$ 
is a set of lines that do not have a non-zero vector in common. The maximum possible cardinality of a line spread is $\frac{2^{s}-1}{3}$ if $s$ is even 
and $\frac{2^{s}-5}{3}$ if $s$ is odd, see e.g.~\cite{beutelspacher}. 

\begin{proposition}
  \label{prop_remove_lines}
  For every integer $s\ge 3$ and every integer $0\le \lambda\le \frac{2^{s-1}-3-2\cdot(-1)^s}{3}$ we have $P(s,2^{s-1}-2\lambda)\le 2^{s}-1-3\lambda$.   
\end{proposition}
\begin{IEEEproof}
  Let $\mathcal{P}$ be the set of $2^s-1$ points in $\mathbb{F}_2^s$ and $\mathcal{R}^i$ the corresponding lists of recovery sets for $1\le i\le s$ of the 
  $s$-dimensional simplex code, see Footnote~\ref{fn_simplex_code}. We will remove $3\lambda$ points from $\mathcal{P}$ and modify the 
  $\mathcal{R}^i$ accordingly. To this end, let $H$ be a hyperplane of $\mathbb{F}_2^s$ not containing any unit vector and $\mathcal{L}$ be a maximal partial 
  line spread of of $H$ of cardinality $\frac{2^{s-1}-1}{3}$ if $s$ is odd and of cardinality $\frac{2^{s-1}-5}{3}$ if $s$ is even. Note that the upper bound 
  for $\lambda$ equals this cardinality. For an arbitrary subset of $\mathcal{L}$ of cardinality $\lambda$ we are remove the corresponding $3\lambda$ points 
  from $\mathcal{P}$ to get the set of points of our final PIR code. We also have to adjust the lists $\mathcal{R}^i$. Since $H$ does not contain any unit 
  vector all elements of the $\mathcal{R}^i$ have cardinality two in the beginning and no recovery set of cardinality two is completely contained in a line 
  from $\mathcal{L}$. So, consider one line that is removed and assume that its points are given by $\{a,b,c\}$. For a fixed but arbitrary $1\le i\le s$ the 
  recovery sets for $e_i$ that contain either $a$, $b$, or $c$ are given by $\{a,a+e_i\}$, $\{b,b+e_i\}$, and $\{c,c+e_i\}.$ Those three recovery sets are destroyed 
  by our operation of removing $\{a,b,c\}$, but we can additionally add the recovery set $\{a+e_i,b+e_i,c+e_i\}$, noting that 
  \begin{eqnarray*}
    &&\left(a+e_i\right)+\left(b+e_i\right)+\left(c+e_i\right)\\ 
    &=&\left(a+b+c\right)+\left(e_i+e_i+e_i\right)=0+e_i=e_i.
  \end{eqnarray*}
  It remains to check that no point $x$ is used in two recovery sets of cardinality three for the same $e_i$ and that we do not remove a point $y$ that is 
  contained in a constructed recovery set of cardinality three. If $x$ is contained in two recovery sets of cardinality three for $e_i$, then $x+e_i$ is 
  contained in two lines of $\mathcal{L}$, which contradicts the disjointness. If $y$ is removed, then $y\in H$. If, additionally, $y$ is contained in a recovery 
  set of cardinality three for $e_i$, then $y+e_i$ is removed, so that $y+e_i\in H$. %% is also contained in $H$. 
  Thus, $e_i\in H$, which is a contradiction.    
\end{IEEEproof}

Note that Proposition~\ref{prop_remove_lines}  e.g.\ improves the best known upper bounds for $P(5,10)$, $P(5,12)$, and $P(5,14)$ compared to \cite[Table I]{fazeli2015private}. 
For $P(5,12)$ we start from the $5$-dimensional simplex code with generator matrix
$$
  \begin{pmatrix}
  00000000 000000000 111111111 111111111\\
  00000000 111111111 000000000 111111111\\
  00011111 000011111 000011111 000011111\\
  01100101 001100101 001100101 001100101\\
  10101011 010101011 010101011 010101011\\
  \end{pmatrix}.
$$
As hyperplane $H$ we can choose the set of $15$ non-zero vectors in $\mathbb{F}_2^5$ that are perpendicular to the all-one vector $(11111)^\top$:
$$
  \begin{smallmatrix}
  000 0000 1111 11111\\
  000 1111 0000 11111\\
  011 0011 0011 00111\\
  110 0101 0101 01010\\
  101 1001 1001 01101\\
  \end{smallmatrix}.
$$ 
Two disjoint lines in $H$ are e.g.\ given by
$$
  \left\{
  \begin{pmatrix}0\\0\\0\\1\\1\\\end{pmatrix},
  \begin{pmatrix}0\\0\\1\\1\\0\\\end{pmatrix},
  \begin{pmatrix}0\\0\\1\\0\\1\\\end{pmatrix}
  \right\}\quad\text{and}\quad
  \left\{
  \begin{pmatrix}0\\1\\0\\0\\1\\\end{pmatrix},
  \begin{pmatrix}1\\0\\0\\0\\1\\\end{pmatrix},
  \begin{pmatrix}1\\1\\0\\0\\0\\\end{pmatrix}
  \right\},
$$
so that we obtain a $5$-dimensional $12$-PIR code with length $25$ and generator matrix
$$
  \begin{pmatrix}
  00000 00000000 11111111 11111111\\
  00000 11111111 00000000 11111111\\
  00111 00011111 00011111 00011111\\
  01001 01100101 01100101 01100101\\
  10011 00101011 00101011 10101011\\
  \end{pmatrix}.
$$

\begin{comment}
While Proposition~\ref{prop_remove_lines} is not applicable directly to conclude $P(4,4)\le 9$, we note that we can obtain the example stated in the appendix 
by starting from a $4$-dimensional simplex code and removing the two lines
$$
  \left\{\begin{pmatrix}1\\1\\0\\0\end{pmatrix},
  \begin{pmatrix}0\\1\\1\\1\end{pmatrix},
  \begin{pmatrix}1\\0\\1\\1\end{pmatrix}\right\}\quad\text{and}\quad
  \left\{\begin{pmatrix}0\\1\\1\\0\end{pmatrix},
  \begin{pmatrix}0\\1\\0\\1\end{pmatrix},
  \begin{pmatrix}0\\0\\1\\1\end{pmatrix}\right\}.
$$
Obviously, the six removed points do not contain a unit vector. However, the construction of the recovery sets has to work differently, since we would have a problem 
for $e_3$ otherwise.
\end{comment}

\begin{corollary}
  For every integer $s\ge 2$ we have $P(s,2^{s-1}-2)=2^s-4$
\end{corollary}
\begin{IEEEproof}
  Since the result is trivial for $s=2$ we assume $s\ge 3$ and consider the lower bound
  $$
    P(s,2^{s-1}-2)\ge \left\lceil\left(2^{s-1}-2\right)\cdot \frac{2^s-1}{2^{s-1}}\right\rceil=2^s-4. 
  $$
  This lower bound is attained by the construction from Proposition~\ref{prop_remove_lines}.
\end{IEEEproof}

\section{Integer linear programming formulations}\label{sec_ILP}
In this section we present an integer linear programming (ILP) formulation for the exact determination of $P(s,k)$. Given the dimension $s$, we set 
$\mathcal{X}=\mathbb{F}_2^s\setminus\{\mathbf{0}\}$. The generator matrix is sufficiently characterized if we know for each element $j\in\mathcal{X}$ the 
integer-valued multiplicity $x_j$ of that column. By $\mathcal{Y}^i$ we denote the set of all minimal recovery sets for $e_i$. For each 
$j\in\mathcal{Y}^i$ we denote by the integer $y_j^i$ the number of times recovery set $j$ for $e_i$ is used. With this, the value of $P(s,k)$ 
is given by
\begin{eqnarray}
  \min && \sum_{j\in\mathcal{X}} x_j\label{ie_target}\\ 
  \sum_{h\in \mathcal{Y}^i\,:\, j\in h} y_h^i\le x_j && \forall j\in \mathcal{X}, \forall 1\le i\le s\label{ie_ILP_linkage}\\
  \sum_{j\in \mathcal{X}\backslash H} x_j \ge k && \forall H\le \mathbb{F}_2^s, \dim(H)=s-1\label{ie_ILP_hyperplane}\\  
   \sum_{h\in\mathcal{Y}^i} y^i_h \ge k && \forall 1\le i\le s\label{ie_ILP_demand}\\
  x_j\in \mathbb N &&\forall j\in \mathcal{X}\label{ilp_x_var}\\ 
    y^i_h\in\mathbb{N} && \forall 1\le i\le s, \forall h\in\mathcal{Y}^i
 \end{eqnarray}  
 Inequality~(\ref{ie_ILP_linkage}) guarantees that the column multiplicities are sufficiently large for the chosen recovery sets. Inequality~(\ref{ie_ILP_demand}) 
 ensures that there are at least $k$ disjoint recovery sets for each $e_i$. Inequality~(\ref{ie_ILP_hyperplane}) implements Lemma~\ref{lemma_hyperplane_argument}.  
 %% models the fact that for each hyperplane $H$ there have to be at least $k$ columns of the generator matrix outside of $H$. 
 In principle, those inequalities are not necessary, but in practice they usually speed up the solution process. 
 Additional lower and upper bounds on $x_j$ can be deduced from Proposition~\ref{prop_lower_bound_multiplicity}, Lemma~\ref{lemma_upper_bound_multiplicity}, respectively. 
 (Tighter bounds can be obtained from more sophisticated coding theoretic arguments, see  
 %%also the discussion on the relation between weights of codewords in the dual code and possible values for the $y$-variables in 
 Section~\ref{sec_dual_distance}.) 

 The problem with that ILP formulation is that it quickly gets too huge to be solved exactly. So, it is applicable for rather small parameters $s$ and $k$ only, where 
 the size of $k$ plays almost no role.

By imposing further restrictions we can use the above ILP formulation as a heuristic to find good codes that eventually attain the known lower bounds or improve 
known constructions from the literature. The restriction to systematic PIR codes can be enforced by $x_{e_i}\ge 1$ for all $1\le i\le s$ and the restriction 
to projective PIR codes can be implemented by $x_j\in\{0,1\}$ for all $j\in\mathcal{X}$. If the dimension $s$ is not too large, then $\# \mathcal{X}$, i.e., the number 
of $x$ variables is still manageable. In order to prevent the combinatorial explosion of $\# \mathcal{Y}^i$ we can restrict ourselves to recovery sets of 
cardinality at most $\lambda$ by modifying the definition of $\mathcal{Y}^i$ accordingly. In our numerical experiments we mostly chose $\lambda=3$ (and $\lambda=4$ 
in some very small cases). The intuition between this heuristic is as follows. We know that the simplex code is optimal and uses recovery sets of cardinalities $1$ and 
$2$ only. For large values of $k$ the coding theoretic lower bound is usually not too far away from the fractional simplex code, i.e., 
the mentioned lower bound $P(s,k)\ge N(s,k)\ge \frac{2^s-1}{2^{s-1}}\cdot k$. So, for a \textit{good} PIR code not too many recovery sets of cardinality larger than $2$ 
can occur. There is some hope that recovery sets of cardinality at most $3$ are sufficient provided $k$ is large enough. The ILP constructions 
in Table~\ref{tab_best_known_bounds} support this hope. For smaller values of $k$ it seems that larger recovery sets are necessary.

Another way to decrease the computational complexity is to prescribe a subgroup of the final automorphism group of the code. In our 
context automorphisms are permutations of the columns and rows of the generator matrix, c.f.\ \cite{freij2018t} where transitive automorphism groups are 
considered. Counting point multiplicities for the columns, as above, this leaves 
row permutations only. So, for some subgroup $H\le S_s$ of the symmetric group on $s$ elements we can require $x_j=x_{\pi j}$ and $y_j^i=y_{\pi j} ^i$ for every $\pi\in H$, 
$1\le i\le j$, and $j\in \mathcal{X}$ or $j\in\mathcal{Y}^i$, respectively. This reduces the number of variables as well as the number of constraints, since several 
of them become identical. If the corresponding substitutions and removals of identical constraints are performed directly this is also known under the 
name Kramer-Mesner approach. As an example we state that prescribing a cyclic $\mathbb{Z}_6$ allowed us to construct an example for $P(6,16)\le 36$ improving 
the previous bound $P(6,16)\le 39$ \cite{vajha2017binary} and prescribing a cyclic $\mathbb{Z}_3$\footnote{Note that for $s=7$ there are two possible 
cycle types of the cyclic group $\mathbb{Z}_3$ in $S_7$ up to conjugation.} allowed us to construct an example for $P(7,16)\le 39$ improving
the previous bound $P(7,16)\le 43$ \cite{vajha2017binary}.

%\medskip

With respect to lower bounds, we remark that it is possible to modify the initial ILP formulation by restricting the possible sizes of recovey sets to 
at most $\lambda$ while still obtaining a lower bound that is valid without this assumption. Given a fix value of $\lambda$ we cannot require 
Inequality~(\ref{ie_ILP_demand}) any more, since we ignore recovery sets that have a cardinality strictly larger than $\lambda$. If we set 
$n=\sum_{j\in\mathcal{X}} x_j$, then we can replace Inequality~(\ref{ie_ILP_demand}) with the following relaxation:
\begin{equation}
  \label{ILP_constraint_new_1}
  n+\sum_{j\in\mathcal{Y}^i} \left(\lambda-\# j\right)\cdot y_j\ge (\lambda+1)\cdot k\quad\forall 1\le i\le s,
\end{equation}
where $\# j$ denotes the size of the recovery set. The idea is simple: $k-\sum_{j\in \mathcal{Y}^i}y_j$ recovery sets for $e_i$ have to be of cardinality 
at least $\lambda+1$ and the number of all recovery sets for $e_i$ cannot be smaller than $n$. So, \textit{choosing} recovery sets of large size has no consequences 
for the $x_j$ directly but imposes lower bounds on $n$, which is a relaxation of the original inequality. In the other direction, Inequality~(\ref{ie_ILP_hyperplane}) 
can be tightened:
\begin{eqnarray}
\label{ie_ILP_hyperplane_improved}
&& \sum_{j\in \mathcal{X}\backslash H} x_j \ge k +\sum_{j\in\mathcal{Y}^i} \# \left(j\cap \mathbb{F}_2^s\backslash H-1\right)y_j\\ %%\quad 
&& \forall 1\le i\le s \forall H\le \mathbb{F}_2^s, \dim(H)=s-1, e_i\notin H,\nonumber 
\end{eqnarray}
where $j\cap \mathbb{F}_2^s\backslash H$ denotes the number of elements of the recovery set $j\in\mathcal{Y}^i$ that are not contained in $H$.
The argument for the hyperplane conditions of Inequality~(\ref{ie_ILP_hyperplane}), see Lemma~\ref{lemma_hyperplane_argument}, was that vectors in $H$ cannot build a 
recovery set for $e_i$ on their own 
for at least one $1\le i\le s$, so that at least one column outside $H$ is needed for each recovery set. Now, if we fix $i$ and we know that some recovery 
sets use more than one column outside of $H$ than the total number of columns outside of $H$ increases, which gives Inequality~(\ref{ie_ILP_hyperplane_improved}).   

We remark, that for $\lambda\le 2$ Inequality~(\ref{ie_ILP_hyperplane_improved}) is the same as Inequality~(\ref{ie_ILP_hyperplane}). Assume the contrary and suppose that for a given 
hyperplane $H$ of $\mathbb{F}_2^s$ and a given unit vector $e_i$ both points of the two-element recovery set $\{a,b\}$ for $e_i$ are not contained in $H$. 
Since the line $\{a,b,a+b+e_i\}$ intersects $H$ in a point, we have $e_i\in H$, which is a contradiction.  

We call the ILP (\ref{ie_target})-(\ref{ie_ILP_hyperplane}), (\ref{ilp_x_var})-(\ref{ie_ILP_hyperplane_improved}) the the lower bound ILP for a given value of $\lambda$. 
We remark that the lower bound ILP for $\lambda=3$ increases the coding theoretic lower bound by $1$ in the cases $(s,k)\in\{(4,3),(4,4),(4,12),(5,10),(5,12),(6,8),$ $(6,12)\}$,  
by $2$ for $(s,k)\in\left\{(5,8),(6,14)\right\}$, and by $4$ for $(s,k)=(6,16)$, cf.~Table~\ref{tab_best_known_bounds}. We remark that the problem 
for $(s,k)=(6,16)$ was too tight to be solved directly. Here we applied symmetry breaking techniques and additional inequalities. More concretely, 
we started by adding $x_{e_1}\ge x_{e_2}\ge\dots x_{e_6}$, $n\le 35$ and maximizing $e_1$. After an upper bound of one was verified we stopped and concluded the additional inequalities 
$x_{e_i}\le 1$ for all $1\le i\le 6$. Similarly, we tried $x_{\mathbf{1}-e_1}\ge x_{\mathbf{1}-e_2}\ge\dots x_{\mathbf{1}-e_6}$, $n\le 35$ and maximized $x_{\mathbf{1}-e_1}$. After 
$14\,028$~seconds and $4\,129\,360$ branch\&bound (B\&B) nodes an upper bound of $1$ was verified, so that $x_{\mathbf{1}-e_i}\le 1$ are valid inequalities for all $1\le i\le 6$ (assuming $n\le 35$). 
With these additional $\le 1$-inequalities we started a last round of symmetry breaking: We introduced the integer variables $s_i$ counting the sum of those $x_p$ where 
point $p$ has a one in the $i$th coordinate. By symmetry we can assume $s_1\ge s_2\ge \dots\ge s_6$. Maximizing $s_1$ with the additional assumption $n\le 35$ was solved after 
$102\,967$~seconds and $24\,767\,382$~B\&B nodes to be infeasible, so that $P(6,16)=36$. We applied the same technique to computationally verify $P(6,14)\ge 32$ in 21\,566~seconds and  
12\,701\,361~B\&B nodes.   

Another interesting instance is:
\begin{proposition}
  \label{prop_non_systematic_is_better}
  Each systematic $6$-dimensional $8$-PIR code of length $n$ satisfies $n\ge 20$, while $P(6,8)=19$.
\end{proposition} 
\begin{IEEEproof}
  We apply the lower bound ILP with $\lambda=3$ while additionally prescribing the use of the vectors from an $s$-dimensional unit 
  matrix. After less than 900~seconds and 125603 B\&B nodes a solution with $n=20$ was proven to be optimal.
\end{IEEEproof}

%******************************************************************************************************************************************************
%******************************************************************************************************************************************************
%******************************************************************************************************************************************************
%******************************************************************************************************************************************************
%******************************************************************************************************************************************************
%******************************************************************************************************************************************************
%******************************************************************************************************************************************************

\section{Lower bounds and the dual minimum distance/dual code(s)}
\label{sec_dual_distance}

%% In this section we present several results on constructions and specific values of $P(s,k)$. We start with the following proposition.
In this section we want to use coding theoretic methods to provide lower bounds for $P(s,k)$. To this end let $C$ be an $[n,s]$ code. The 
corresponding dual code $C^\perp$ is the $[n,n-s]$ code whose codewords are those that are perpendicular to all codewords in $C$. 
By $d^\perp$ we denote the minimum distance of $\C^\perp$, which is also called the dual minimum distance. 

\begin{lemma}
  \label{prop_d_perp_lower_bound}
  Let $C$ be a linear $[n,s,d]$ code with minimum dual distance $d^\perp$ and generator matrix $G$.
  \begin{itemize}
    \item[(a)] If $R$ and $R'$ are two different recovery sets for the same symbol $i$ in $G$, then $|R|+ |R'|\ge d^\perp$.
    \item[(b)] If $G$ is a $k$-PIR generator matrix with $k\ge 2$, then $n\ge k\lceil d^\perp/2\rceil-1$ if $d^\perp$ is odd and $n\ge k\lceil d^\perp/2\rceil$ 
               if $d^\perp$ is even.
    \item[(c)] If $G$ is a $k$-PIR generator matrix that contains a unit vector $e_i$, then 
               $n\ge 1+(k-1)(d^\perp-1)$.           
  \end{itemize}
\end{lemma}
\begin{IEEEproof}
  Since $\sum_{j\in R} G^j = e_i=\sum_{j\in R'} G^j$, we have $\sum_{j\in (R\cup R')\backslash(R\cap R')} G^j =0$, i.e., $(R\cup R')\backslash(R\cap R')$ is the support 
  of a dual codeword, which gives part~(a). Next, we consider an arbitrary unit vector $e_i$ and let $m$ denote the cardinality of the smallest recovery set for $e_i$. 
  From (a) we conclude that every other recovery set has cardinality at least $d^\perp-m$, so that $n\ge (k-1)(d^\perp-m)+m$. The special case $m=1$ corresponds 
  to (c). For part~(b) we can argue as follows. If $m\ge \lceil d^\perp/2\rceil$, then $n\ge k\lceil d^\perp/2\rceil$, so that we assume $m\le \lceil d^\perp\rceil-1$ and 
  conclude
  $$
    n\ge (k-1)(d^\perp-m)+m \ge (k-2)\lceil d^\perp/2\rceil+d^\perp.
  $$  
\end{IEEEproof}

\begin{proposition}
  \label{prop_reed_muller_no_PIR}
  For each integer $s\ge 4$ we have $P(s,2^{s-2})\ge 2^{s-1}+1$.
\end{proposition}
\begin{IEEEproof}
  It is well known that $N(s,2^{s-2})=2^{s-1}$ with the unique solution being the first order Reed-Muller code, i.e., in geometric terms, %%the affine part of an $s$-dimensional simplex (
  all points of $\mathbb{F}_2^s$ except those in a distinguished hyperplane, see e.g.\ Lemma~\ref{lemma_reed_muller_unique} for a short self-contained proof. As no multiple 
  points or lines (sets of three collinear points) are contained, the dual minimal distance $d^\perp$ is at 
  least $4$ (indeed it is $4$). Let $G$ be a generator matrix that is a $2^{s-2}$-PIR code. If $G$ contains a unit vector, 
  then Lemma~(\ref{prop_d_perp_lower_bound}).(c) gives $n\ge 1+\left(2^{s-2}-1\right)\cdot 3$, which is a contradiction for $s\ge 4$. Thus, $G$ does not contain any 
  unit vector and every recovery set has cardinality at least $2$. Since $n=2k$ every recovery set has cardinality exactly $2$.
  
  In order to obtain a contradiction we now prove the following statement by induction on $1\le j\le s-1$. For each $1\le j\le s-1$ there exist vectors 
  $x_1,\dots, x_l\in \mathbb{F}_2^s$, where $l=2^{s-1-j}$, such that the columns of $G$ are given by $\left\{ x_h+\langle e_1,\dots,e_j\rangle \mid 1\le h\le l\right\}$, 
  where we slightly abuse notation. By $x_h+\langle e_1,\dots,e_j\rangle$ we abbreviate the list of $2^j$ vectors contained in the affine $\mathbb{F}_2$-vector space 
  $x_h+\langle e_1,\dots,e_j\rangle$.  
  
  For the induction start we remark that the $2^{s-1}$ columns are partitioned into pairs $\left\{x_h,x_h+e_1\right\}$ corresponding to the recovery sets of 
  $e_1$. For the induction step we assume that the columns of $G$ are partitioned into $l=2^{s-1-j}$ sets, which we call blocks, of the form 
  $x_h+\langle e_1,\dots,e_j\rangle$. Now we are considering the recovery sets of $e_{j+1}$. Let $R=\{a,b\}$ a recovery set of $e_{j+1}$ that was not considered before. 
  Note that $a$ and $b$ have to be contained in different blocks since $a+b=e_{j+1}$. W.l.o.g. let $a$ be contained in the first block 
  $x_1+\langle e_1,\dots, e_j\rangle$ and $b$ in the second block $x_2+\langle e_1,\dots, e_j\rangle$, so that we can reparameterize to 
  $a+\langle e_1,\dots, e_j\rangle$ and $b+\langle e_1,\dots, e_j\rangle$. Since $a+b=e_{j+1}$ the union of the two blocks can be described by 
  $\mathcal{B}=a+\langle e_1,\dots, e_j,e_{j+1}\rangle$. Note that this new block $\mathcal{B}$ contains $2^j$ recovery sets for $e_{j+1}$ of cardinality $2$. In principle 
  those recovery sets do not need to coincide with those from which we started. However, we can perform the following swaps. Let $\{b_1,c_1\}$ be an so far 
  unconsidered recovery set for $e_{j+1}$ of cardinality $2$ with $b_1\in\mathcal{B}$ and $c_1\notin\mathcal{B}$. Let $b_2\in \mathcal{B}$ with $b_1+b_2=e_{j+1}$ 
  and $c_2\notin \mathcal{B}$ with $c_2+b_2=e_{j+1}$. Instead of the recovery pairs $\{b_1,c_1\}$ and $\{b_2,c_2\}$ we swap to the recovery pairs 
  $\{b_1,b_2\}\subseteq \mathcal{B}$ and $\{c_1,c_2\}$. Thus, we can assume that all nodes of $\mathcal{B}$ pair within $\mathcal{B}$. Going  on with 
  another unconsidered recovery pair gives us a new block each time so that the induction step is proven.
  
  For $e_s$ we use the structural information that the columns of $G$ can be described as $x+\langle e_1,\dots,e_{s-1}\rangle$ for some vector $x\in\mathbb{F}_2^s$. 
  Thus there can be no recovery set of cardinality two for $e_s$.  
  
Since the first order Reed-Muller code is excluded, we have $P(s,2^{s-2})\ge N(s,2^{s-2})+1=2^{s-1}+1$. 
\end{IEEEproof}
%%Note that in the proof of Proposition~\ref{prop_reed_muller_no_PIR} we use the label representation for recovery sets and not the vector representation. 

We remark that the uniqueness of the first order Reed-Muller code is not needed in the above proof. It is sufficient to have the information that any 
length optimal code with dimension $s$ and minimum distance $2^{s-2}$ satisfies $d^\perp\ge 4$, which can e.g.\ be concluded from the MacWilliams equations.

\begin{comment}
Additionally, we remark that it should be possible to rewrite the latter part of the proof using the number $a_4^*$ of codewords of the dual code of weight $4$. 
It can be easily figured out that for an $[2^{s-1},s,2^{s-2}]$ code we have
$$
  a_4^*=\frac{2^{s-3}-6\cdot 4^{s-3}+8\cdot 8^{s-3}}{3}.
$$
Any pair of recovery sets of cardinality $2$ for $e_i$ gives rise to a dual codeword of weight $4$. However, the same dual codeword of weight $4$ can correspond to 
a pair of recovery sets of cardinality $2$ for $e_i$ as well as for $e_j$ with $i\neq j$. So we need to analyze how often this can happen. In general, the number of dual 
codewords of small weight tells us something about the possible sizes of recovery sets. 
\end{comment}

Another application of Lemma~\ref{prop_d_perp_lower_bound} is to use the uniqueness of the binary extended Golay code with parameters $[24,12,8]$, see 
\cite{pless1968uniqueness}\footnote{The code is even unique within the class of non-linear codes, see \cite{delsarte1975unrestricted}.} Since the code is 
self-dual, we have $d^\perp=8$ so that part~(b) implies that any generator matrix of the binary extended Golay code cannot be a $7$-PIR code. Since 
$N(12,8)=24$ this implies $P(12,8)\ge 25$.

As a further relation between the minimum dual distance $d^\perp$ and PIR codes we note that the lower bound (\ref{ie_lower_bound_vardy_rao}) was proven in 
\cite{wootters2016linear} using the dual code and especially the minimum dual distance.

\begin{lemma}
\label{lemma_reed_muller_unique}
Let $s\ge 1$ and $\ell\ge 0$ be integers and $C$ be a linear $\left[2^{s-1}+\ell\left(2^{s}-1\right),s,(2\ell+1)2^{s-2}\right]$ code. \red{If $\mathcal{P}$ denotes the corresponding multiset of points, then the multiplicity of every point in $\mathbb{F}_2^s$ with respect to $\mathcal{P}$ is either $\ell$ or $\ell+1$}. Moreover, the $2^{s-1}-1$ points with multiplicity $\ell$ form a hyperplane in $\mathbb{F}_2^s$.
\end{lemma}
\begin{IEEEproof}
  Assume that there exists a point $p$ with multiplicity larger or equal to $\ell+2$. W.l.o.g.\ we assume that $p=e_i$ for some $1\le i\le s$, so that shortening gives an 
  $\left[\le(2\ell+1)\cdot 2^{s-1}-2\ell-2,s-1,(2\ell+1)2^{s-2}\right]$ code, which is a contradiction since each $[n',s-1,(2\ell+1)2^{s-2}]$ code satisfies 
  $n'\ge (2\ell-1)\cdot\left(2^{s-1}-1\right)=(2\ell+1)\cdot 2^{s-1}-2\ell-1$. Now consider the complementary multiset of points $\mathcal{P}'$ where the multiplicity of 
  each point of $\mathbb{F}_2^s$ is given by $\ell+1$ minus its original multiplicity with respect to $\mathcal{P}$. Counting points gives that $|\mathcal{P}'| =2^{s-1}-1$. 
  Let $H$ be an arbitrary hyperplane of $\mathbb{F}_2^s$. Due to Inequality~(\ref{ie_hyperplane_argument}) in Lemma~\ref{lemma_hyperplane_argument} we have 
  $|\left(\mathcal{P}'\cap H\right)|\ge 2^{s-2}-1$ for every hyperplane. Now we are using linear combinations of the left hand and right hand sides of the standard equations.  
  Using the abbreviation $x=2^{s-2}-1$, $x(2x+1)$ times Equation~(\ref{se1}) minus $3x$ times Equation~(\ref{se2}) plus Equation~(\ref{se3}) gives 
  $$\sum_{i\ge 0} (2x+1-i)(x-i) h_i=(x+1)y_2.$$
  Due to Inequality~(\ref{ie_hyperplane_argument}) we have $h_i=0$ for $i<x$ and since the number of points if $2x+1$, we have $h_i=0$ for $i>2x+1$. The coefficient 
  $(2x+1-i)(x-i)$ of $h_i$ is zero for $i\in\{x,2x+1\}$ and strictly negative for all $x<i<2x+1$. Since $h_i\ge 0$ for all integer $i$ the left hand side is at most zero. 
  From $x\ge 0$ and $y_2\ge 0$ we conclude that the right hand side is at least zero, so that both side have to be equal to zero. This directly implies $y_2=0$ and $h_i=0$ 
  for all $x<i<2x+1$. From Equation~(\ref{se1}) and Equation~(\ref{se2}) we then conclude $h_x=2^{s-2}$ and $h_{2x+1}=1$. $y_2=0$ tells us that the point multiplicity with 
  respect to $\mathcal{P}'$ is at most $1$, so that the point multiplicity with respect to $\mathcal{P}$ is at least $\ell$. From $h_{2x+1}=1$ we read of that there is exactly 
  one hyperplane $H$ whose $2x+1=2^{s-1}-1$ points form the set $\mathcal{P}'$, so that the stated result follows. 
\end{IEEEproof}

We remark that a more complicated proof has been given for example in \cite{chen2009notes}. However, the result should be well-known for several decades. %(unfortunately I do not have an older citation at hand).

\begin{lemma}
  \label{lemma_simplex_reduction}
  Let $C$ be an $s$-dimensional binary $k$-PIR code of length $n$ that contains every non-zero vector of $\mathbb{F}_2^s$ at least once as a column of a generator matrix, 
  then $n\ge P(s,k-2^{s-2})+2^{s-1}-1$.
\end{lemma}
\begin{IEEEproof}
  Let $\mathcal{R}^i$ be corresponding recovery sets. We will now show that we can modify the recovery sets so that they contain the recovery sets of the $s$-dimensional 
  simplex code as a subset. First of all, we assume that all recovery sets in $\mathcal{R}^i$ are minimal. Especially, we have that $\{e_i\}$ is contained in 
  $\mathcal{R}^i$. Due to symmetry we only consider the modification of $\mathcal{R}^1$.  
  %% If $e_1$ is contained in a recovery set of cardinality strictly larger than $1$, then we remove the other vectors from that recovery set. 
  For every vector $x\in\mathbb{F}_2^s\backslash \mathbf{0}$ with first coordinate equal to zero we have the 
  recovery set $\{x,x+e_1\}$ in the simplex code. If that recovery set is contained in $\mathcal{R}^i$ that's fine. Otherwise $x$ is contained in a recovery set $A$ with $|A|\ge 3$ and 
  $x+e_1$ is contained in a recovery set $B$ with $|B|\ge 3$. We replace the recovery sets $A$ and $B$ by $\{x,x+e_1\}$ and $A\cup B\backslash \{x,x+e_1\}$ (considered as a 
  multiset union or with removed duplicates).    
\end{IEEEproof}

From Proposition~\ref{prop_reed_muller_no_PIR}, Lemma~\ref{lemma_reed_muller_unique} and Lemma~\ref{lemma_simplex_reduction} we iteratively conclude:
\begin{corollary}
  \label{cor_lower_bound_reed_muller}
  For each integer $s\ge 4$ and each integer $\ell\ge 0$ we have $P(s,2^{s-2}+\ell 2^{s-1})\ge  \ell\left(2^{s}-1\right)+2^{s-1}+1$.
\end{corollary}
An example is $P(4,12)\ge 24$, which can indeed be attained and improves the coding theoretic lower bound $N(4,12)=23$ by one. We note that we only 
need the information that every non-zero point of $\mathbb{F}_2^s$ is taken at least once, i.e., $y_2=0$ for the complementary multiset of points. 

Another example, which is a bit more involved, is $P(5,8)\ge 18$. %%Again, the coding theoretic lower bound $N(5,8)=17$ is improved by one.
\green{The coding theoretic lower bound $N(5,8)=16$ is improved by two}.
\begin{lemma}
  \label{lemma_M_5_8_projective}
  If $P(5,8)=17$, then the corresponding PIR code is projective and has weight distribution \red{$0^1 8^{14} 9^{16} 16^1$}.
\end{lemma}
\begin{IEEEproof}
  Let $P$ be a column with multiplicity $m$. Shortening then gives a $[17-m,4,8]$ code $C'$, which implies $m\le 2$. So, we assume $m=2$ 
  and note that the unique $[15,4,8]$ code is the $4$-dimensional simplex code. I.e., the columns of a generator matrix $G'$ of $C'$ consist 
  of all non-zero vectors of $\mathbb{F}_2^4$. Now consider any lengthened $[17,5,8]$ code. The two new columns can be contained in at most 
  two different recovery sets for $e_5$. Recovery sets for $e_5$ that consist solely of some of the first $15$ columns of $G'$ have cardinality 
  at least three, since $C'$ has dual minimum distance $3$, i.e., no two columns of $G'$ sum to zero. Thus, $2\cdot 1+6\cdot 3>17$ gives a 
  contradiction, so that $m=1$ and the code has to be projective.
  
  Finally, we note that there are exactly four $[17,5,\ge 8]$ codes. Only one of these is projective and has the stated weight 
  distribution.\footnote{All exhaustive lists of binary linear codes have been enumerated with the software package \texttt{LinCode}, 
  see \cite{LinCode}.\label{fn_lincode}} 
\end{IEEEproof}

\begin{comment}
We remark that there are exactly four $[17,5,\ge 8]$ codes. Their weight and dual weight distributions are given 
as follows:
\begin{itemize}
  \item $0^1 8^{25} 12^6$; $0^1 2^1 3^{15} 4^{85} 5^{231} 6^{349}\dots$
  \item $0^1 8^{22} 10^8 16^1$; $0^1 2^1 3^{7} 4^{133} 5^{119} 6^{469}\dots$
  \item $0^1 8^{14} 9^{16} 16^1$; $0^1 3^{8} 4^{140} 5^{112} 6^{448}\dots$
  \item $0^1 8^{15} 9^{15} 17^1$; $0^1 2^1 4^{175} 6^{721}\dots$
\end{itemize}
It is easy to exclude the last case as a $8$-PIR code. If the generator matrix contains $e_i$ for some $1\le i\le s$, then $\mathcal{R}^i$ 
does not contain recovery sets of cardinality $2$, since $a_3^\star=0$. Since $e_i$ can be used at most twice we obtain the contradiction 
$2\cdot 1+(8-2)\cdot 3=20>17$. So, for every $1\le i\le s$ all except at most one of the eight recovery sets hat cardinality two and one might have 
cardinality $3$. Since $a_2^\star=2$ we have exactly one doubled column of $G$, which we call $x$. Obviously $x$ cannot be contained in two different 
recovery sets for some $e_i$ since otherwise we would have two doubled columns. Since $x$ also cannot be contained twice in a (reduced) recovery set, 
for all $1\le i\le s$ we have exactly $7$ recovery sets of cardinality $2$ and one cardinality set of cardinality $3$ that has to contain vector $x$ 
exactly once. Since this gives a dual codeword of weight $5$, we observe a contradiction.
\end{comment}

The unique code determined in Lemma~\ref{lemma_M_5_8_projective} has a dual weight distribution starting with 
$0^1 3^{8} 4^{140} 5^{112} 6^{448}\dots$ and can be generated by
$$
  \begin{pmatrix}
  11111110000010000\\
  11110001110001000\\
  11001101101000100\\
  00111101100100010\\
  10101011011100001\\
  \end{pmatrix}.
$$
\begin{comment}
For each $e_i$ there can be at most one recovery set of cardinality $1$, so that there have to be at least five recovery sets of cardinality $2$. In 
total at most one column of the generator matrix can be a unit vector since two unit vectors would generate $2\cdot 5-1>8$ dual codewords of 
weight $3$ ($e_i$ plus a cardinality two recovery set, where we can have at most one coincidence). So, the generator matrix is non-systematic at 
the very least. For at least four unit vectors we have exactly seven recovery sets of cardinality $2$ and one of cardinality $3$. 
\end{comment}

The weight distribution of dual codes can be used even more directly than in the proof of Lemma~\ref{lemma_M_5_8_projective}. Assume that we have an 
$s$-dimension $k$-PIR code of length $n$ with generator matrix $G$. Let $G'$ denote the matrix that arises if we remove the $i$th row of $G$. 
\begin{comment}
%%\begin{example}
\label{example_generator_matrix_recovery_sets}
An example for a generator matrix attaining $P(4,4)=9$ and $i=4$ is 
given by: 
$$
  G=\begin{pmatrix}
    1 0 0 0 1 1 1 1 1\\ 
    0 1 0 0 0 1 0 1 1\\ 
    0 0 1 0 1 1 0 0 1\\ 
    0 0 0 1 0 0 1 1 1\\
  \end{pmatrix}
  \quad
  G'=\begin{pmatrix}
    1 0 0 0 1 1 1 1 1\\ 
    0 1 0 0 0 1 0 1 1\\ 
    0 0 1 0 1 1 0 0 1\\ 
  \end{pmatrix}.
$$
\end{comment}
The recovery sets of cardinality $w$ for $e_i$ in $G$ correspond to dual codewords in $G'$ of weight exactly $w$. Obviously, this is not a bijection, since we 
completely ignore the entries in the $i$th row of $G$. 

\begin{lemma}\label{lemma_expurgation}
\red{Let $G$ is the generator matrix of an $s$-dimensional $k$-PIR code of length $n$. If the cardinality vector of $\mathcal{R}^i$, where $1\le i\le s$ is arbitrary, is given by $1^{m_1} 2^{m_2} 3^{m_3}\dots$ (clearly $\sum_j m_j=k$), then there exists a matrix $G'$ that is the generator matrix of an $(s-1)$-dimensional $k$-PIR code of length $n-m_1$ such that there exist $k-m_1$ disjoint dual codewords with weight distribution $2^{m_2} 3^{m_3}\dots$.}%%(Disjoint \textit{recovery sets} for the zero vector.)  
\end{lemma}
\begin{IEEEproof}
  Apply expurgation, i.e., remove the $i$th row from $G$.
\end{IEEEproof}

\begin{theorem}
  \label{thm_5_8}
  $$P(5,8)=18$$
\end{theorem}
\begin{IEEEproof}
\green{Due to Corollary~\ref{cor_lower_bound_reed_muller} we only have to consider length $n=17$.}  
We apply Lemma~\ref{lemma_expurgation} and enumerate the \green{$[17,4,\ge 8]$} codes $C_i$. There are exactly 23 of them. However, 
we can use more information of a putative $5$-dimensional $8$-PIR code of length $17$. The possible cardinality vectors of one list 
$\mathcal{R}^i$ are
$$
  1^1 2^5 3^2, 1^1 2^6 3^1, 1^1 2^6 4^1, 1^1 2^7, 2^7 3^1, 
$$
i.e., in any case we have at least five recovery sets of cardinality $2$. For the codes $C_i$ this translates to the requirement that 
in a generator matrix there have to be at least five disjoint pairs of identical columns and at least $7$ disjoint pairs of identical columns   
if the effective length is $17$. This leaves the following four codes with generator matrices
\begin{eqnarray*}
&&
\left(\begin{smallmatrix}
1111111000001000\\
1111000111000100\\
1100110110100010\\
0011110110010001\\
\end{smallmatrix}\right),
\left(\begin{smallmatrix}
11111110000001000\\
11110001110000100\\
11001101101100010\\
00111101100010001\\
\end{smallmatrix}\right),\\ 
&&
\left(\begin{smallmatrix}
11111110000001000\\
11110001110000100\\
11001101101100010\\
11110000001110001\\
\end{smallmatrix}\right),
\left(\begin{smallmatrix}
11111110000001000\\
11110001110000100\\
11110000001110010\\
10001101101100001\\
\end{smallmatrix}\right),
\end{eqnarray*}
see Footnote~\ref{fn_lincode}. 

The last one contains a column with multiplicity $3$. Since by adding an additional row to the generator matrix the multiplicity of each point can decrease by a factor 
of at most $2$, this contradicts Lemma~\ref{lemma_M_5_8_projective}. 

The first one is a doubled Reed-Muller code with dual weight distribution $0^1 2^8 4^{252} 6^{952}\dots$. 
Of course the code itself is a $8$-PIR code. No recovery set of cardinality one can be used, since there is no dual codeword of weight $3$ and at 
least one recovery set of cardinality $2$ has to be used for each $e_i$. Thus, every cardinality set has exactly cardinality $2$. So, everything 
could be partitioned into two half's and we would obtain two $4$-dimensional $4$-PIR codes of length $8$ each, which do not exist. 

The second code has weight distribution $0^1 8^7 9^4 11^4$, so that it clearly cannot be augmented to a code with weight distribution $0^1 8^{14} 9^{16} 16^1$ due to the codeword of 
weight $11$. 

So there remains the third code with weight distribution $0^1 8^6 9^8 16^1$. %% and dual weight distribution $0^1 2^8 3^{16} 4^{252} 5^{224} 6^{952}\dots$.
Thus, a $5$-dimensional $8$-PIR code has a generator matrix without any unit vector, since expurgation would otherwise give a $4$-dimensional code of length strictly less than $17$. 
So, the cardinality distribution for every $\mathcal{R}^i$ is $2^73^1$ and all rows of the generator matrix have a weight of exactly $8$. (In a recovery set of cardinality $2$ 
for $e_i$ there is exactly one $1$ in coordinate $i$ and in a recovery set of cardinality $3$ there are either $1$ or $3$ ones in coordinate $i$. Since there is no codeword of 
weight $10$ in the code the stated observation follows.) So, the weight of any row of the generator matrix is divisible by $8$, the sum of any two rows has a weight divisible 
by $4$, and the sum of any three different rows has a weight divisible by $2$. %%(Obvious with the divisible code background.) 
Thus, the number of codewords of weight $9$ is at most ${5\choose 4}+{5\choose 5}=6<16$, which is a contradiction.
\end{IEEEproof}

\begin{comment}
  \textbf{ToDo:} The above argument is like poking around in the dark. There are several ad hoc arguments that partially increase our knowledge, which is then put 
  together in a chaotic way. Either a shortest path through the arguments should be found or, the better option, the underlying techniques should be extracted 
  and generalized. (If it is really worth the effort, the computer generated code listings may be replaced by coding theoretic hand computations, which, of course, 
  won't decrease the length of the argumentation.)
\end{comment}

\section{Bounds and exact values of PIR codes}\label{sec:results}
%The exact values of $P(s,1)$ and $P(s,2)$ are well-known for all integers $s\ge 1$. \begin{proposition} \label{prop_demand_2_exact} For all integers $s\ge 1$ we have $P(s,1)=s$ and $P(s,2)=s+1$. \end{proposition} \begin{IEEEproof} Looking at a systematic generator matrix we see that an $s$-dimensional linear code has obviously length at least $s$. Taking the $s$-dimensional unit matrix we conclude $N(s,1)=1$ so that $P(s,1)$, since every systematic symbol can be recovered from the corresponding unit vector. The result for $k=2$ is then implied by Equation~(\ref{ie_parity_check}). \end{IEEEproof} We remark that in geometrical terms the codes attaining $P(s,2)=s+1$ are the so-called $s$-dimensional projective bases. For each $e_i$ the two recovery sets have  cardinality $1$ and $s$.

Lemma~\ref{lemma_simplex_reduction} has another important consequence.
\begin{proposition}
  \label{prop_lower_bound_multiplicity}
  For each positive integer $\ell\ge P(s,k)-2k$ we have
  $$
    P\!\left(s,k+2^{s-1}\cdot \ell\right)=P\!\left(s,k+2^{s-1}\cdot (\ell-1)\right)+2^{s}-1.
  $$
\end{proposition}
\begin{IEEEproof}
  Since $P(s,2^{s-1})\le 2^s-1$, %% (actually, we have equality), 
  we obviously have $P\!\left(s,k+2^{s-1}\cdot \ell\right)\le P\!\left(s,k+2^{s-1}\cdot (\ell-1)\right)+2^{s}-1$. 
  Now let $G$ be a generator matrix of a matching PIR code attaining length $P\!\left(s,k+2^{s-1}\cdot \ell\right)$. If every non-zero vector of $\mathbb{F}_2^s$ occurs as a column 
  of $G$, then Lemma~\ref{lemma_simplex_reduction} gives $P\!\left(s,k+2^{s-1}\cdot \ell\right)\ge P\!\left(s,k+2^{s-1}\cdot (\ell-1)\right)+2^{s}-1$. Thus, it remains to 
  assume the existence of a non-zero point $P\in\mathbb{F}_2^s$ with multiplicity zero. By $x_j$ we denote the number of occurrences of vector $j\in\mathbb{F}_2^s\backslash\mathbf{0}$ 
  as column vector of $G$. From Inequality~(\ref{ie_hyperplane_argument}) we conclude $\sum_{j\notin H} x_j\ge k'$ for every hyperplane $H$ of $\mathbb{F}_2^s$, where 
  $k'=k+2^{s-1}\cdot \ell$. Summing over all $2^{s-1}$ hyperplanes that do not contain $P$ gives  
  $$    \sum_{H\le \mathbb{F}_2^s\,:\,\dim(H)=s-1, P\notin H} \sum_{j\notin H} x_j\ge 2^{s-1}\cdot k'.   $$ 
  The coefficient of $x_P$ on the left hand side is $2^{s-1}$ and for any other non-zero point $Q\neq P$ the coefficient of $x_Q$ on the left hand side 
  is given by 
  $$
    \frac{2^{s-1}\cdot \left(2^{s-1}-1\right)}{2^s-2}=2^{s-2},
  $$
  so that $n\ge 2k'$, where $n=\sum_j x_j$ (using $x_P=0$). Thus,
  $$
    n\ge 2k'=\left(2k+l\right)+\left(2^s-1\right)\cdot \ell
  $$
  Since $P\!\left(s,k+2^{s-1}\cdot \ell\right)\le P(s,k)+\left(2^s-1\right)\cdot \ell$, stated result follows.
\end{IEEEproof}

As an example we use $P(4,4)=9$ to conclude $P(4,12)=24$, or more generally $P(4,4+8\ell)=9+15\ell$ for every integer $\ell\ge 0$, \brown{c.f.\ 
Proposition~\ref{prop_dimension_4_exact}}. We remark that the condition $\ell\ge P(s,k)-2k$ in Proposition~\ref{prop_lower_bound_multiplicity} can be 
replaced by $\ell\ge \mu-2k$, where $\mu$ is some arbitrary upper bound for $P(s,k)$.

\begin{corollary}
  \label{cor_baumert_generalization}
  For every fix integer $s\ge 1$ the determination of the function $P(s,\cdot)$ is a finite problem.
\end{corollary}
\begin{IEEEproof}
  \brown{Choose an integer $\ell$ such that $\ell\ge P(s,k)-2k$ for all $1\le k\le 2^{s-1}$. Due to Proposition~\ref{prop_lower_bound_multiplicity}, the 
  values $P(s,k)$ for $k\le \ell\cdot 2^{s-1}$ completely determine the function $P(s,\cdot)$.}
\end{IEEEproof}

While the above example $P(4,4+8\ell)=9+15\ell$ that $P(s,k)$ does not need to tend to the coding theoretic lower bound $N(s,k)$, which exactly approaches 
the Griesmer bound, for sufficiently large values of $k$, Corollary~\ref{cor_baumert_generalization} generalizes the result from \cite{baumert1972note} for 
the Griesmer bound to PIR codes.

In a similar style, as in the proof of Proposition~\ref{prop_lower_bound_multiplicity}, we can state an easy to evaluate lower bound on the code size 
if some point has a large multiplicity:

\begin{lemma}\label{lemma_upper_bound_multiplicity}
\red{Let $C$ be an $[n,s,d]$ code where one column of a generator matrix has multiplicity $m$. Then, $n\ge \frac{2^{s-1}-1}{2^{s-2}}\cdot d+m$.}
\end{lemma} 
\begin{IEEEproof}
  By $x_j$ we denote the number of columns of a given generator matrix $G$ of $C$ that are equal to a non-zero vector $j\in\mathbb{F}_2^s$. Let $P$ be non-zero 
  vector with multiplicity $m$. Summing Inequality~(\ref{ie_hyperplane_argument}) over all hyperplanes that contain $P$, we obtain
  $$
    \sum_{H\le \mathbb{F}_2^s\,:\,\dim(H)=s-1, P\in H} \sum_{j\notin H} x_j\ge \left(2^{s-1}-1\right)\cdot d.
  $$
  The coefficient of $x_P$ on the left hand side is $0$ and for any other non-zero point $Q\neq P$ the coefficient of $x_Q$ on the left hand side 
  is given by 
  $$
    \frac{2^{s-1}\cdot \left(2^{s-1}-1\right)}{2^s-2}=2^{s-2},
  $$
  so that
  $$
    n-m\ge \frac{2^{s-1}-1}{2^{s-2}}\cdot d.
  $$
\end{IEEEproof}

Of course the condition multiplicity $m$ in Lemma~\ref{lemma_upper_bound_multiplicity} can be replaced by condition multiplicity at least $m$ and we can 
also reformulate the inequality to $m\le n-\frac{2^{s-1}-1}{2^{s-2}}$, where $d$ can be replaced by $k$ in the context of PIR codes.

\begin{comment}
If the dimension $s$ is at most $3$, the exact values of $P(s,k)$ are also well known.
\begin{proposition}
  \label{prop_dimension_at_most_3_exact}
  \begin{enumerate}
    \item[(a)] For every integer $k\ge 1$ we have $P(1,k)=k$.
    \item[(b)] For every integer $k\ge 1$ we have $P(2,k)=\lceil 3k/2\rceil$.
    \item[(c)] For every even integer $k\ge 2$ we have $P(3,k)=\lceil 7k/4\rceil$.
  \end{enumerate}  
\end{proposition}
\end{comment}
%\begin{IEEEproof} The lower bounds are all implied by the coding theoretic lower bound $P(s,k)\le N(s,k)$. The $1$- and the $2$-dimensional simplex codes give $P(1,1)\le 1$ and $P(2,2)\le 3$, so that the stated result for $s\le 2$ follows by replication and the parity-check transform. For $s=3$, the $3$-dimensional simplex code gives $P(3,4)\le 7$. For $P(3,2)\le 4$ we can use the projective base of dimension $3$, see Proposition~\ref{prop_demand_2_exact}. \end{IEEEproof}

Following up Corollary~\ref{cor_baumert_generalization}, we solve \brown{two} further instances:
%%As a new result we obtained:
\begin{proposition}
  \label{prop_dimension_4_exact}
  We have $P(4,2)=5$, $P(4,4)=9$, $P(4,6)=12$, $P(4,8)=15$, and for all even $k>8$ we have $P(4,k)=P(4,k-8\tau)+15\tau$, where $\tau=\lfloor(k-1)/8\rfloor$.
\end{proposition}
\begin{IEEEproof}
  For $k\in\{2,4,6,8\}$ the corresponding \brown{upper} bounds were known before and can be easily verified using the ILP approach.   
  The other constructions are then obtained by combinations with a suitable number of $4$-dimensional 
  simplex codes. Except for $P(4,4+8\ell)\ge 9+15\ell$ the coding theoretic lower bound is attained. For the latter lower bound 
  see Corollary~\ref{cor_lower_bound_reed_muller}. (According to Proposition~\ref{prop_lower_bound_multiplicity} it is sufficient to prove $P(4,4)\ge 9$, which 
  can be done using Inequality~(\ref{ie_lower_bound_vardy_rao}).) %% or the ILP lower bound with $\lambda=3$, which also is easily computable for $P(4,12)$.)   
\end{IEEEproof}

\begin{proposition}
  \label{prop_dimension_5_exact}
  We have \red{$P(5,2)=6$, $P(5,4)=10$}, $P(5,6)=14$, $P(5,8)=18$, $P(5,10)=22$, $P(5,12)=25$, $P(5,14)=28$, \red{$P(5,16)=31$} and for all even $k>16$ we have 
  $P(5,k)=P(5,k-16\tau)+31\tau$, where $\tau=\lfloor(k-1)/16\rfloor$.
\end{proposition}
\begin{IEEEproof}
  For $k\in \{2,4,6,16\}$ the \brown{upper} bounds were known before and for $k\in\{8,10,12,14\}$ the \brown{upper} bounds have been found using the ILP approach, see 
  the corresponding generator matrices: 
  \begin{eqnarray*}
  &&   
   \left(\begin{smallmatrix}
 1 0 0 0 0 1 0 1 0 1 0 1 0 1 0 1 0 1 0 1 0 1\\ 
 0 1 0 0 0 1 0 0 1 1 0 0 1 1 0 0 1 1 0 0 1 1\\ 
 0 0 1 0 0 0 1 0 0 0 1 1 1 1 0 0 0 0 1 1 1 1\\ 
 0 0 0 1 0 0 1 0 0 0 0 0 0 0 1 1 1 1 1 1 1 1\\ 
 0 0 0 0 1 0 0 1 1 1 1 1 1 1 1 1 1 1 1 1 1 1\\ 
  \end{smallmatrix}\right), 
   \left(\begin{smallmatrix}
 1 0 0 0 0 1 1 1 0 1 0 1 0 1 0 1 0 1 0 0 1 0 1 0 1\\ 
 0 1 0 0 0 0 1 0 1 1 0 0 1 1 1 1 0 0 1 0 0 1 1 0 1\\ 
 0 0 1 0 0 1 1 0 0 0 1 1 1 1 0 0 1 1 1 0 0 0 0 1 1\\ 
 0 0 0 1 0 0 0 1 1 1 1 1 1 1 0 0 0 0 0 1 1 1 1 1 1\\ 
 0 0 0 0 1 0 0 0 0 0 0 0 0 0 1 1 1 1 1 1 1 1 1 1 1\\ 
  \end{smallmatrix}\right),\\ 
  &&
   \left(\begin{smallmatrix}
 1 0 0 0 0 1 1 0 1 1 0 0 1 0 1 0 1 0 1 0 1 0 1 1 0 1 0 1\\ 
 0 1 0 0 0 1 0 1 1 0 1 0 0 1 1 1 1 0 0 1 1 0 0 1 0 0 1 1\\ 
 0 0 1 0 0 0 1 1 1 0 0 1 1 1 1 0 0 1 1 1 1 0 0 0 1 1 1 1\\ 
 0 0 0 1 0 0 0 0 0 1 1 1 1 1 1 0 0 0 0 0 0 1 1 1 1 1 1 1\\ 
 0 0 0 0 1 0 0 0 0 0 0 0 0 0 0 1 1 1 1 1 1 1 1 1 1 1 1 1\\
 \end{smallmatrix}\right),
 \left(\begin{smallmatrix}
     1 0 0 0 0 1 1 0 1 1 0 0 1 0 0 1 0 1\\ 
     0 1 0 0 0 0 1 1 1 0 1 0 0 1 0 0 1 1\\ 
     0 0 1 0 0 0 1 0 0 1 1 0 0 0 1 1 1 1\\ 
     0 0 0 1 0 1 1 0 0 0 0 1 1 1 1 1 1 1\\ 
     0 0 0 0 1 0 0 1 1 1 1 1 1 1 1 1 1 1\\ 
  \end{smallmatrix}\right).
\end{eqnarray*}
  The other constructions are then obtained by combinations 
   with a suitable number of $5$-dimensional simplex codes. Except for $P(5,8+16\ell)\ge 18+31\ell$, $P(5,10+16\ell)\ge 22+31\ell$, and $P(5,12+16\ell)\ge 25+31\ell$  the coding 
  theoretic lower bound is attained. By Proposition~\ref{prop_lower_bound_multiplicity} it remains 
  to prove $P(5,8)\ge 18$, $P(5,24)\ge 49$, $P(5,10)\ge 22$, $P(5,26)\ge 53$, and $P(5,12)\ge 25$. To this end we have utilized the ILP lower bound 
  for $\lambda=3$, see Subsection~\ref{sec_ILP}. (Each computation took just a few seconds.)
\end{IEEEproof}
\brown{We remark that also Proposition~\ref{prop_remove_lines} gives matching constructions for $k\in \{10,12,14\}$.}

%As mentioned in the introduction we have: \begin{proposition} For every positive integer $s$ we have $P(s,2^{s-1})=2^{s}-1$.\end{proposition}

%%Maybe the result can also be adjusted for batch codes.

%%\section{Results}
%%\label{sec_results}

In Table~\ref{tab_best_known_bounds} we state the best known bounds for $P(s,k)$ for small parameters. Improvements over the existing literature are printed 
in bold. We use the following letters to point to the method from which the bound was obtained.
\begin{itemize}
  \item[c] The coding theoretic lower bound $P(s,k)\ge N(s,k)$, see Inequality~(\ref{ie_coding_theoretic_lower_bound}).
  \item[r] The lower bound of Rao and Vardy, see Inequality~(\ref{ie_lower_bound_vardy_rao}).
  \item[e] The exact values of $P(s,2)$ and $P(s,k)$ for $s\le 3$ are well-known, see Theorem~\ref{thm_known_results}.
  \item[a] Two PIR codes can be easily combined to a new one, see Theorem~\ref{thm_known_results}.10)-11).
  \item[S] $s$-dimensional simplex code.
  \item[I] Constructive ILP approach, see Section~\ref{sec_ILP}. (Generator matrices can be obtained from the authors or the arXiv version of this paper.) 
  \item[i] ILP lower bound with $\lambda=3$, see Section~\ref{sec_ILP}.
  \item[L] Lengthening, see Proposition~\ref{prop_lengthening} and the subsequent algorithm, see also \cite{lin2017lengthening}.
  \item[R] Binary shortened projective Reed-Muller codes, see \cite{vajha2017binary}.
\end{itemize}

\begin{table*}[htp]
\begin{center}
\caption{Best known bounds for $P(s,k)$ for small parameters.}  \label{tab_best_known_bounds}
    \begin{tabular}{rrrrrrrrrr}
      \hline
      $s/k$ & 2 & 3 & 4 & 6 & 8 & 10 & 12 & 14 & 16  \\ 
      \hline
       1 &  $^c{2}^e$ &  $^r\overline{3}^e$ &  $^r{4}^e$ &  $^c{6}^e$ &  $^c{8}^e$ & $^c{10}^e$ & $^c{12}^e$ & $^c{14}^e$ & $^c{16}^e$ \\
       2 &  $^c{3}^e$ &  $^r\overline{5}^e$ &  $^r{6}^e$ &  $^c{9}^e$ & $^c{12}^e$ & $^c{15}^e$ & $^c{18}^e$ & $^c{21}^e$ & $^c{24}^e$ \\ 
       3 &  $^c{4}^e$ &  $^r\overline{6}^e$ &  $^r{7}^e$ & $^c{11}^e$ & $^c{14}^e$ & $^c{18}^e$ & $^c{21}^e$ & $^c{25}^e$ & $^c{28}^e$ \\ 
       4 &  $^c{5}^e$ &  $^r$8$^I$ &  $^r$9$^I$ & $^c{12}^I$ & $^c{15}^S$ & $^c{20}^a$ & $^i$\textbf{24}$^a$ & $^c{27}^a$ & $^c{30}^a$ \\ 
       5 &  $^c{6}^e$ &  $^r$9$^I$ & $^r$10$^I$ & $^c$14$^I$ & $^i$\textbf{18}$^I$ & $^i\mathbf{{22}}^I$ & $^i\mathbf{{25}}^I$ & $^c\mathbf{{28}}^I$ & $^c{31}^S$ \\ 
       6 &  $^c{7}^e$ & $^r$10$^I$ & $^r$11$^I$ & $^c$15$^I$ & $^i$\textbf{19}$^I$ & $^c$\textbf{23}$^I$ & $^i$\textbf{27}$^I$ & $^i$\textbf{32}$^I$ & $^i$\textbf{36}$^I$ \\ 
       7 &  $^c{8}^e$ & $^r$12$^{R}$ & $^r$13$^{R}$ & $^c$\textbf{16}$^I$ & $^c$19--\textbf{21}$^I$ & $^c$24--\textbf{26}$^I$ & $^c$27--\textbf{29}$^I$ & $^c$32--\textbf{34}$^I$ & $^c$35--\textbf{39}$^I$ \\ 
       8 &  $^c{9}^e$ & $^r$13$^{R}$ & $^r$14$^{R}$ & $^c$17--18$^{L}$ & $^c$20--\textbf{23}$^{L}$ & $^c$26--\textbf{27} & $^c$29--\textbf{33}$^I$ & $^c$33--\textbf{38}$^I$ & $^c$36--\textbf{42}$^I$ \\ 
       9 & $^c{10}^e$ & $^r$14$^{R}$ & $^r$15$^{R}$ & $^c$18--20$^{L}$ & $^c$21--25$^{R}$ & $^c$27--\textbf{28}$^I$ & $^c$30--\textbf{37}$^I$ & $^c$35--\textbf{40}$^I$ & $^c$38--\textbf{45}$^I$ \\ 
      10 & $^c{11}^e$ & $^r$15$^{R}$ & $^r$16$^{R}$ & $^c$20--21$^{L}$ & $^c$22--26$^{R}$ & $^c$28--\textbf{31}$^L$ & $^c$31--\textbf{40}$^I$ & $^c$36--\textbf{45}$^I$ & $^c$40--50$^I$ \\
      \hline 
    \end{tabular}
  \end{center}
\end{table*}

\begin{comment}
\textbf{ToDo:}\\ 
\begin{itemize}
  \item I do not know the generator matrices for the upper bounds for $P(9,6)$, $P(10,6)$ (from \cite{lin2017lengthening} obtained by lengthening).
  \item I do not know the generator matrices for the upper bounds for $P(9,8)$, $P(10,8)$ (from \cite{vajha2017binary} obtained from 
        shortened projective Reed-Muller codes).
  \item I do not know the generator matrices for the upper bounds for $P(7,4)$, $P(8,4)$, $P(9,4)$, $P(10,4)$ (from \cite{vajha2017binary} obtained from 
        shortened projective Reed-Muller codes; for dimensions $7\le s\le 9$ these codes were also known before).      
\end{itemize}  
\end{comment}
 
%% Compared to \cite[Table I]{fazeli2015private} the improvements of Table~\ref{tab_best_known_bounds} are quite significant 
%% in several cases, i.e., $P(9,16)\le 60$ was improved to $P(9,16)\le 45$. (The result $P(9,8)\le 25$ from \cite{lin2017lengthening} 
%% implies $P(9,16)\le 50$, in \cite{vajha2017binary} the authors obtained $P(9,16)\le 46$.)

We remark that Table~\ref{tab_best_known_bounds} contains improvements of the lower bound $P(s,k)\ge N(s,k)$ in the cases
$(s,k)\in\{(4,3),(4,4),(4,12),(5,8),(5,10),(5,12),(6,8),(6,12),$ $(6,14)\}$. All of these can be obtained with the 
ILP lower bound described in Section~\ref{sec_ILP}, for some we state a coding theoretic proof in Section~\ref{sec_dual_distance}.

Looking at the differences between the coding theoretic lower bound $N(s,k)$ and the best known lower bound for $P(s,k)$ it seems that $P(s,2^{s-2})$ 
is an instance where the optimal codes have a large length compared to the coding theoretic lower bound. 
\begin{conjecture}
  $$\lim_{s\to\infty} P(s,k)-N(s,k)=\infty,$$
  where $k=2^{s-2}$.
\end{conjecture}
Prescribing the cyclic groups $\mathbb{Z}_s$ as a subgroup of the automorphism group we have obtained the following upper bounds for these instances, where 
the coding theoretic lower bound is attained by the first order Reed-Muller codes:
\begin{itemize}
  \item $P(6,16)\le 36$; cardinalities of the recovery sets: $1^1 2^{10} 3^5$
  \item $P(7,32)\le 71$; cardinalities of the recovery sets: $1^1 2^{23} 3^8$
  \item $P(8,64)\le 142$; cardinalities of the recovery sets: $1^1 2^{48} 3^{15}$
  \item $P(9,128)\le 282$; cardinalities of the recovery sets: $1^1 2^{100} 3^{27}$
\end{itemize}
As a general construction we may use Proposition~\ref{prop_remove_lines}:
\begin{corollary}
  For $s\ge 5$ we have
  $$
    P\!\left(s,2^{s-2}\right)\le 2^s-1-3\cdot 2^{s-3}=5\cdot 2^{s-3}-1.
  $$
\end{corollary}
Note that this gives $P(5,8)\le 19$, $P(6,16)\le 39$ (which is also an improvement to \cite[Table I]{fazeli2015private}), $P(7,32)\le 79$, $P(8,64)\le 159$, 
and $P(9,128)\le 319$. In general we have $P(s,2^{s-2})\ge N(s,2^{s-2})= 2^{s-1}=4\cdot 2^{s-3}$.

%\bibliography{PIR}

\begin{thebibliography}{10}
\providecommand{\url}[1]{#1}
\csname url@rmstyle\endcsname
\providecommand{\newblock}{\relax}
\providecommand{\bibinfo}[2]{#2}
\providecommand\BIBentrySTDinterwordspacing{\spaceskip=0pt\relax}
\providecommand\BIBentryALTinterwordstretchfactor{4}
\providecommand\BIBentryALTinterwordspacing{\spaceskip=\fontdimen2\font plus
\BIBentryALTinterwordstretchfactor\fontdimen3\font minus
  \fontdimen4\font\relax}
\providecommand\BIBforeignlanguage[2]{{%
\expandafter\ifx\csname l@#1\endcsname\relax
\typeout{** WARNING: IEEEtran.bst: No hyphenation pattern has been}%
\typeout{** loaded for the language `#1'. Using the pattern for}%
\typeout{** the default language instead.}%
\else
\language=\csname l@#1\endcsname
\fi
#2}}

\bibitem{B16}
J. Bornholt, R. Lopez, D. M. Carmean, L. Ceze, G. Seelig, and K. Strauss, 
``A DNA-based archival storage system," 
\emph{ASPLOS}, pp. 637--649, Atlanta, GA, Apr. 2016.



\bibitem{baumert1972note}
L.~Baumert and R.~McEliece.
\newblock A note on the {G}riesmer bound.
\newblock {\em IEEE Transactions on Information Theory}, 19(2):134--135, 1973.

\bibitem{beutelspacher}
A.~Beutelspacher.
\newblock Partial spreads in finite projective spaces and partial designs. 
\newblock {\em Mathematische Zeitschrift}, 145(3), 211--229, 1975.

\bibitem{bouyukhev2000smallest}
I.~Bouyukhev, D.~B. Jaffe, and V.~Vavrek.
\newblock The smallest length of eight-dimensional binary linear codes with
  prescribed minimum distance.
\newblock {\em IEEE Transactions on Information Theory}, 46(4):1539--1544,
  2000.

\bibitem{carvalho2019projective}
C.~Carvalho, X.~Ram{\'\i}rez-Mondrag{\'o}n, V.~G. Neumann, and
  H.~Tapia-Recillas.
\newblock Projective {R}eed--{M}uller type codes on higher dimensional scrolls.
\newblock {\em Designs, Codes and Cryptography}, pages 1--16, 2019.

\bibitem{chen2009notes}
Y.~Chen and H.~Vinck.
\newblock Notes on {R}eed-{M}uller codes.
\newblock {\em arXiv preprint 0901.2062}, 2009.

\bibitem{costello1982error}
D.~J. Costello and S.~Lin.
\newblock {\em Error Control Coding: Fundamentals and Applications}.
\newblock Pearson, Prentice Hall, Upper Saddle River, NJ, USA, 2nd edition,
  2004.

\bibitem{delsarte1975unrestricted}
P.~Delsarte and J.-M. Goethals.
\newblock Unrestricted codes with the golay parameters are unique.
\newblock {\em Discrete Mathematics}, 12(3):211--224, 1975.

\bibitem{7282977}
A.~Fazeli, A.~Vardy, and E.~Yaakobi.
\newblock Codes for distributed {P}{I}{R} with optimal storage overhead.
\newblock In {\em 2015 IEEE International Symposium on Information Theory
  (ISIT)}, pages 2852--2856, June 2015.

\bibitem{fazeli2015private}
A.~Fazeli, A.~Vardy, and E.~Yaakobi.
\newblock {P}{I}{R} with low storage overhead: {C}oding
  instead of replication.
\newblock {\em arXiv preprint 1505.06241}, 2015.

\bibitem{FGW17}
{S.L.\,Frank-Fischer, V.\,Guruswami, and M.\,Wootters},
``Locality via partially lifted codes," arXiv:1704.08627, Apr. 2017.

\bibitem{freij2018t}
R.~Freij-Hollanti, O.~W. Gnilke, C.~Hollanti, A.-L. Horlemann-Trautmann,
  D.~Karpuk, and I.~Kubjas.
\newblock t-private information retrieval schemes using transitive codes.
\newblock {\em IEEE Transactions on Information Theory}, 2018.

\bibitem{Grassl:codetables}
M.~Grassl.
\newblock Bounds on the minimum distance of linear codes and quantum codes.
\newblock Online available at \url{http://www.codetables.de}, 2007.
\newblock Accessed on 2019-01-24.

\bibitem{griesmer1960bound}
J.~H. Griesmer.
\newblock A bound for error-correcting codes.
\newblock {\em IBM Journal of Research and Development}, 4(5):532--542, 1960.

\bibitem{LinCode}
S.~Kurz.
\newblock LinCode -- computer classification of linear codes.
\newblock {\em arXiv preprint 1912.09357}, 2019.

\bibitem{LC04}
S. Lin and D. J. Costello, 
\emph{Error Control Coding}, 
Prentice Hall, 2004. 

\bibitem{lin2017lengthening}
H.-Y. Lin and E.~Rosnes.
\newblock Lengthening and extending binary private information retrieval codes.
\newblock {\em arXiv preprint 1707.03495}, 2017.

\bibitem{lin1980class}
S.~Lin and G.~Markowsky.
\newblock On a class of one-step majority-logic decodable cyclic codes.
\newblock {\em IBM Journal of Research and Development}, 24(1):56--63, 1980.

\bibitem{lipmaa2015linear}
H.~Lipmaa and V.~Skachek.
\newblock Linear batch codes.
\newblock In {\em Coding Theory and Applications}, pages 245--253. Springer,
  2015.

\bibitem{Massey}
J.L. Massey, \emph{Threshold Decoding}, MIT Press, 1963.

\bibitem{PHO13}
L. Pamies-Juarez, H. D. Hollmann, and F. Oggier, ``Locally repairable codes with multiple repair alternatives," in \emph{Proc. IEEE Int. Symp. Inf. Theory}, pp. 892--896, Istanbul, Turkey, Jul. 2013. 

\bibitem{pless1968uniqueness}
V.~Pless.
\newblock On the uniqueness of the {G}olay codes.
\newblock {\em Journal of Combinatorial theory}, 5(3):215--228, 1968.

\bibitem{ramkumar2018determining}
V.~Ramkumar, M.~Vajha, and P.~V. Kumar.
\newblock Determining the generalized hamming weight hierarchy of the binary
  projective {R}eed-{M}uller code.
\newblock {\em arXiv preprint 1806.02028}, 2018.

\bibitem{rao2016lower}
S.~Rao and A.~Vardy.
\newblock Lower bound on the redundancy of {P}{I}{R} codes.
\newblock {\em arXiv preprint 1605.01869}, 2016.

\bibitem{RPDV14}
A. Rawat, D. Papailiopoulos, A. Dimakis, and S. Vishwanath, ``Locality and availability in distributed storage," in \emph{Proc. IEEE Int. Symp. Inf. Theory}, pp. 681--685, Honolulu, HI, Jun. 2014.

\bibitem{skachek2018batch}
V.~Skachek.
\newblock Batch and {P}{I}{R} codes and their connections to locally repairable
  codes.
\newblock In {\em Network Coding and Subspace Designs}, pages 427--442.
  Springer, 2018.

\bibitem{vajha2017binary}
M.~Vajha, V.~Ramkumar, and R.~V. Kumar.
\newblock Binary, shortened projective {R}eed {M}uller codes for coded private
  information \green{retrieval}.
\newblock In {\em Information Theory (ISIT), 2017 IEEE International Symposium
  on}, pages 2648--2652. IEEE, 2017.

\bibitem{WZL15}
A. Wang, Z. Zhang , and M. Liu, 
``Achieving arbitrary locality and availability in binary codes,"
in \emph{Proc. IEEE Int. Symp. on Inf. Theory}, pp.\,1866--1870, Hong Kong, Jun. 2015.

\bibitem{wootters2016linear}
M.~Wootters.
\newblock Linear codes with disjoint repair groups.
\newblock {\em unpublished mansucript, February}, 2016.


\end{thebibliography}
%\bibdata{PIR}
%\bibliographystyle{abbrv}

%\vspace{-2ex}

\end{document}